\documentclass[
reprint, amsmath,amssymb, aps,
 pre, longbibliography,
]{revtex4-1}

\newcommand{\partDeriv}[2]{\frac{\partial #1}{\partial #2}}

\usepackage{graphicx}
\usepackage{dcolumn}
\usepackage{bm}
\usepackage{hyperref}
\usepackage{float}

\begin{document}

\title{On The Diffusion of Sticky Particles in 1-D}

\author{Joshua DM Hellier}
 \email{J.D.M.Hellier@sms.ed.ac.uk}
\author{Graeme J Ackland}
 \email{G.J.Ackland@ed.ac.uk}
\affiliation{
 SUPA, School of Physics and Astronomy, University of Edinburgh, Mayfield Road, Edinburgh EH9 3JZ, United Kingdom
}

\date{\today}

\begin{abstract}
The 1D Ising model is the simplest Hamiltonian-based model in statistical mechanics. The sim-
plest interacting particle process is the Symmetric Exclusion Process (SEP), a 1D lattice gas of
particles that hop symmetrically and cannot overlap. Combining the two gives a model for sticky
particle diffusion, SPM, which is described here. SPM dynamics are based on SEP with short-range
interaction, allowing flow due to non-equilibrium boundary conditions. We prove that SPM is also
a detailed-balance respecting, particle-conserving, Monte Carlo (MC) description of the Ising model. Neither the Ising model nor SEP have a phase transition in 1D, but the SPM exhibits a non-equilibrium
transition from a diffusing to a blocked state as stickiness increases.  This transition manifests in peaks in the MC density fluctuation, a change in the dependency of flow-rate on stickiness, and odd structure in the eigenspectrum of the transition rate matrix.
We derive and solve a fully non-linear, analytic, mean-field solution, which has a crossover from a positive to a negative diffusion constant where the MC shows the transition.
The negative diffusion constant in fact indicates a breakdown of the mean-field approximation, with close to zero flow and breaking into a two-phase mixture, and thus the mean field theory successfully predicts its own demise.
We also present an analytic solution for the flow via direct analysis of the transition rate matrix.  The
simplicity of the model suggest a wide range of possible applications.

\end{abstract}

\maketitle

 \begin{figure} \vspace{1em}
\caption{\label{fig:rates} White circles indicate particles, dark circles indicate empty sites (vacancies). Particles randomly move into adjacent vacancies with rate $1$ (having rescaled time for notational convenience), unless there is a particle behind the position they're moving from, in which case they move with rate $\lambda$; the state of the site next to the position the particle is moving into is irrelevant.}
    \includegraphics[width=\linewidth]{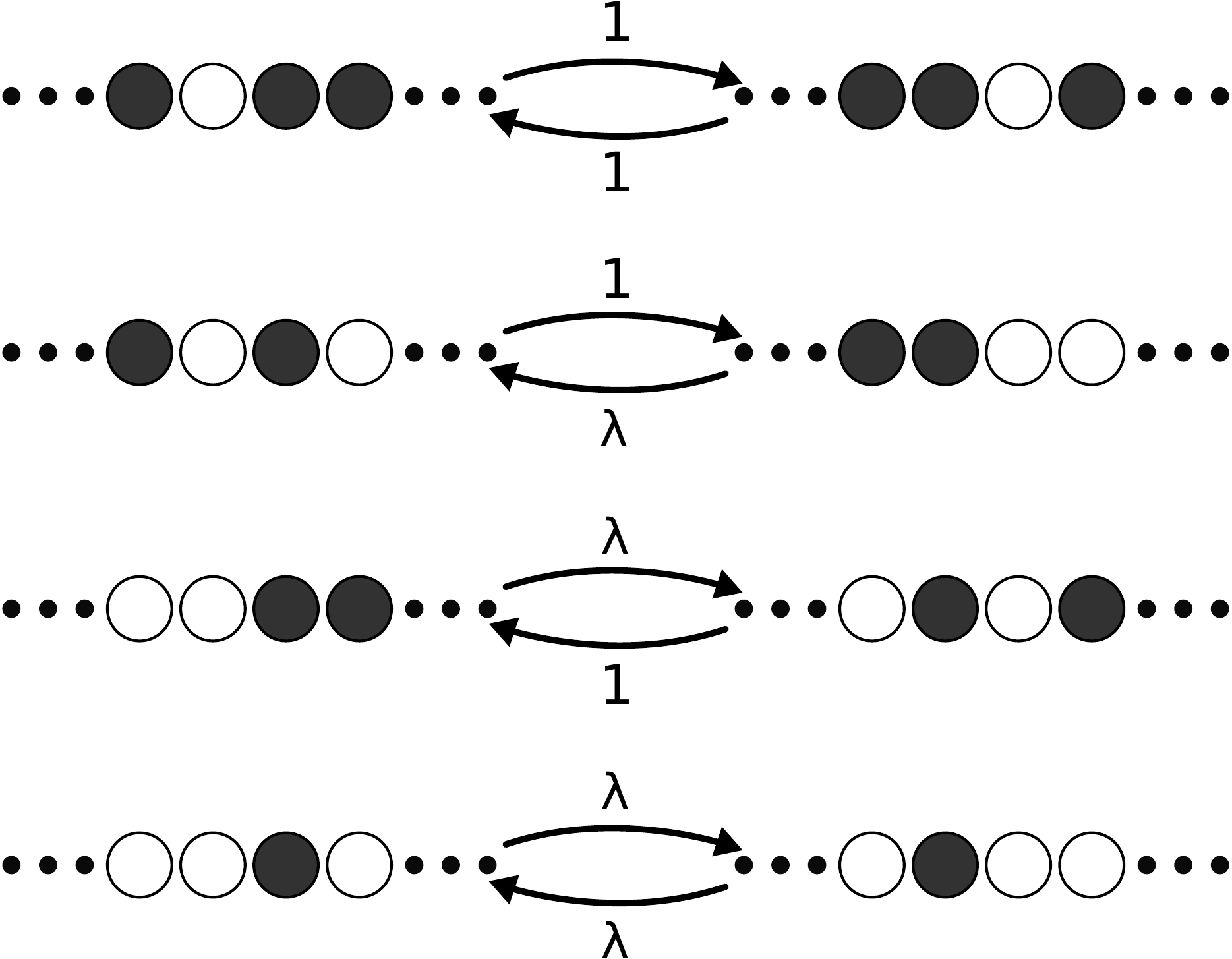}
    \end{figure}

\section{Introduction}

Lattice gases are a ubiquitous tool for modeling complex systems from
biology to traffic~\cite{1742-5468-2011-07-P07007, Mobilia2007,
  tegner2015high, zhu2012atomic, DealGrove1965, MottCabrera1949,
  Buzzaccaro2007}.  Analytically solvable cases involve
non-interacting or excluding particles~\cite{ladd1988application,
  liggett1985interacting, BenNaim1999, Shandarin1989, Frachebourg1999,
  Frachebourg2000}, but in any real system of interest the moving
objects interact. Many models tackle the situation where the diffusing
objects interact with the substrate, but despite clear
application-relevance there is surprisingly little work considering
interactions between the moving particles themselves.  One reason for
this is that the interactions introduce nonlinearities in analytical
models, which makes them challenging to solve, at least outside of
limits in which they can be linearized. This is unfortunate because it
is precisely these nonlinearities which introduce interesting
behaviors such as discontinuities at interfaces or
diffusion instability~\cite{Obukhovsky2017, Gorokhova2010}.

Another feature of previous models is that the flow is driven by
either asymmetry in the dynamics (e.g. the Asymmetric Simple Exclusion Process, ASEP) or an external field
which permeates the system (e.g. the Katz-Lebowitz-Spohn model, KLS).  In either case, the particles
always see a local asymmetry.  However in many systems the flow is
driven by a pressure or chemical potential difference applied at the
boundaries, so that any local asymmetry arises from self-organization.
This is the situation that is addressed here.

In this paper we will investigate what we believe to be the simplest interacting one-dimensional exclusion model, the
``Sticky Particle Model'' or SPM, which is specified in Fig.~\ref{fig:rates}. We
will explore the impact this has on particle behavior, in particular
when observed in the large-scale limit. The model will be fully defined in Section~\ref{sec:modelDefn}, and we will show that it obeys detailed balance away from boundaries. We will then discuss a method to solve the system exactly for small systems in Section~\ref{sec:transMatrix}, and Monte-Carlo techniques for larger systems in Section~\ref{sec:simulations}. In Section~\ref{sec:mftPred} we will attempt to understand our results by means of a simple MFT approximation.  One might contrast this
approach (making a simple microscopic model and trying to learn about large-scale properties) with approaches such as the KPZ
equation~\cite{PhysRevLett.56.889, PhysRevA.38.4271, Sasamoto2010}
(where one analyses the extreme large-scale dynamics using
universality classes).
Throughout the simulations and analytics, we aim to address three kinds of questions: how does stickiness affect the flow, how does the
particle current depend on the driving force, and what determines the density of particles in the system?

\section{The sticky particle model} \label{sec:modelDefn}

The SPM is an excluded-particle model in which adjacent particles
separate with rate $\lambda$ and single particles move at rate $1$.

It differs from the symmetric exclusion
process~\cite{sugden2007dynamically, Kollmann2003, Lin2005,
  Hegde2014,Krapivsky2014, Imamura2017} in that particles ``unstick''
with rate $\lambda$ instead of their normal hopping rate, $1$.  Low
$\lambda$ corresponds to sticky particles, high $\lambda$ to repelling
particles.
It could be regarded as a version of the KLS model~\cite{Katz1984,
  Zia2010, Kafri2003} in 1-dimension without an applied field, which
is itself similar to the dynamics used to analyze the Ising model by
Kawasaki~\cite{PhysRev.145.224}; it is a member of the class of models considered in
\cite{spohn1983}.  The KLS model has a field that
introduces an asymmetry which in turn drives a flow.  It is tacitly assumed
that this field is required for the nontrivial flow behavior seen in
the model, so the simpler symmetric model has received less attention.

\subsection{Detailed Balance Proof} 
The SPM is intended to study flow, so it is {\bf defined} by the
hopping rates.  In Fig.~\ref{fig:rates} we
show all the possible transitions which may occur between local
configurations. 

Assume that the system is now on a ring, with $L$
lattice sites and $N$ particles.  Let us label possible system
configurations by $\xi$ and let the number of adjacencies (or
``bonds'') between particles be $b(\xi)$. Now for our ansatz, assume
that the probability of the system being in state $\xi$ is
$\lambda^{-b(\xi)}$.  In the top and bottom diagrams of
Fig.~\ref{fig:rates} we can see that the number of bonds on
both sides is the same, as are the transition rates back and forth;
thus our ansatz holds for these states, as it predicts the
probabilities of the left and right configurations are the same. The
middle two diagrams are equivalent; in the upper diagram a
bond is formed going left to right and then broken going right to
left, so the probability of being in the left state is $\lambda$ times
that of being in the right state. This is again in agreement with the
detailed balance criterion. As these are the only types of transition
that may occur on a ring, we have proven that the closed system obeys
detailed balance, with an energy proportional to $b(\xi) \ln {\lambda}$.

\subsection{Equivalence to Ising Model in Equilibrium}

Since the model obeys detailed balance, there must be an associated
Hamiltonian.  This is simply proportional to the number of particles stuck together:
\begin{equation}
 H = q \sum_i p_ip_{i+1} = q b(\xi)
\end{equation}
where $p_i$ is 1 for a particle or zero for an empty site.  A simple
transformation $s_i = 2p_i-1$ shows this to be equivalent to the Ising
Hamiltonian
\begin{equation}
 H = q \sum_i \left[ \frac{1}{4} s_i s_{i+1} + \left( \frac{1}{2} s_i + \frac{1}{4} \right) \right].
\end{equation}
The contribution from the final term is constant, and because only hopping moves are allowed
$\sum s_i$ is constant\cite{PhysRev.145.224}. 
Therefore in equilibrium the SPM samples the 1D
Ising model with fixed magnetization, a fact which was used to
validate the codes. Compatibility between the detailed balance condition, the standard Boltzmann distribution at temperature $T$ and our defined rates requires that $q = k_B T\ln{\lambda} $.

\subsection{Nonequilibrium Calculations}

Of course, the real point of our research into the SPM is to see how it behaves out of equilibrium. We drive a system of length $L$ out of equilibrium by attaching it to
reservoirs with densities $\rho_0$ and $\rho_L$ at the $x=0$ and $x=L$ ends respectively,
with a single fixed value for $\lambda$ throughout the system.
We have computed the main properties of this system (principally the current and the average density of particles) by using three methods: numerical analysis of the transition rate matrix (TRM) for a small system, Monte Carlo (MC) methods for somewhat larger systems and a mean-field
approximation for large systems in the continuum limit. These methods and their results are discussed in Sections~\ref{sec:transMatrix}, \ref{sec:simulations} and \ref{sec:mftPred}. respectively.

\section{Transition Rate Matrix Calculations}
\label{sec:transMatrix}
The SPM, just like the KLS model and ASEP, can exhibit flow. In this paper we will be principally concerned with the computation of the properties of the steady state of an SPM system of $L$ sites
connected to (unequal) particle reservoirs at either end.

It is possible to ``analytically'' solve the SPM on a finite domain with fixed boundary densities by analyzing the transition rate matrix which represents that system. For a system of length $L+4$ this is a sparse matrix with dimension 
$2^{L+4} \times 2^{L+4}$.
The non-zero matrix elements relating to bulk motion are $\lambda$ or $1$ while those relating to the boundary conditions are given by Eqs.~\ref{eq:appRate} and~\ref{eq:disRate}. 
The matrix represents a finite irreducible continuous time Markov process, so it is guaranteed to have a unique zero eigenvalue,
whose eigenvector represents the long-term steady state flow. All other eigenvalues have negative real part and represent fluctuations.
There are a pair of additional sites at each boundary
representing the particle reservoirs, hence the overall system size being recorded as $L + 4$. In these boundary layers particles rapidly pop in and out of existence,
with rates such that the mean occupation should be the desired boundary density. The choice of how to achieve this is not unique, but in our computations the rate at which particles appear in empty spaces
at the boundary is 
\begin{equation}
\label{eq:appRate}
C_0 (1+\lambda)\sqrt{\lambda\frac{\rho_0}{1-\rho_0}},
\end{equation}
and the rate at which particles disappear is
\begin{equation}
\label{eq:disRate}
C_0 (1+\lambda)\sqrt{\lambda\frac{1-\rho_0}{\rho_0}}.
\end{equation}

In isolation, the above rates would ensure that the average occupation was always $\rho_0$. We have an equivalent set of creation and annihilation rates at the other boundary attempting to maintain the density at $\rho_L$. In our overall SPM system, we attempt to ensure that all of these rates are larger than, or at the very least not smaller than,
the other rates in the system ($1$ and $\lambda$); this should ensure that the occupation of the boundary sites is always approximately $\rho_0$, regardless of what is happening internally.
Thus, $C_0$ is chosen to be quite large, and the $\lambda$-dependence exists primarily in order to ensure that $\lambda$ does not dominate the boundary rates when it becomes large.
It is worth noting that with the addition of these boundary conditions, the model as a whole no longer obeys detailed balance, even though the bulk still does in terms of
local behavior.

Once we have built the correct transition rate matrix, we then use the
Python routine \texttt{scipy.sparse.linalg.eigs} to find the
eigenvector and associated eigenvalue with the most positive real
part; this eigenvalue
should be zero, up to numerical error, and the corresponding
eigenvector describes the system's
stationary state.  The stationary state contains most of the key
information we desire about the system's behavior during steady flow,
including the current and particle density profile, which may be
accessed by the application of occupation and current operators
(constructed along with the transition rate matrix). In this steady
state, we are guaranteed to have a uniform current flowing through the
interior of the system; this is implied by the uniqueness of the
steady state and the homogeneity of the internal dynamics.  The next
least-negative eigenpairs dominate the approach to equilibrium, and
determine the timescale over which it is achieved.

\begin{figure}[h!]
\vspace{0em}
\begin{center}
    \includegraphics[width=1\linewidth]{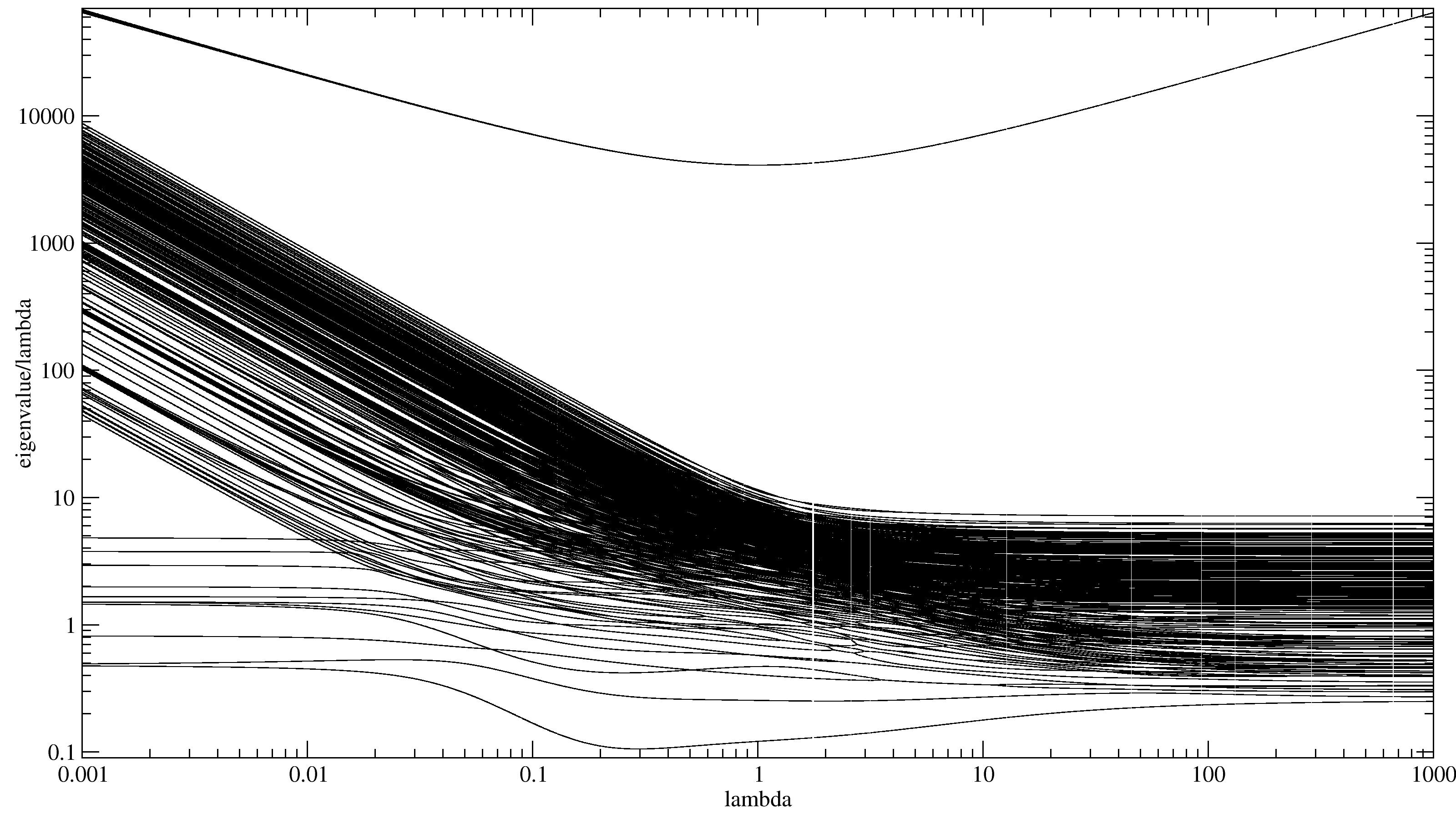}
\end{center}
    \vspace{-0em}
\caption{\label{fig:eigenspec} The moduli of the $1024$ eigenvalues of
  the transition rate matrix with smallest negative real part, as a
  function of $\lambda$.
  Moduli are shown divided by $\lambda$ to emphasize the 
  $\mathcal{O}(\lambda)$ and  $\mathcal{O}(\lambda^2)$  trends.
  There is always a single zero eigenvalue These smallest eigenvalues
  correspond to the decay rates of the dominant transient states in
  the system, implying that the time taken for a simulated system to
  relax to its equilibrium state scales as $\lambda^{-1}$. This helps
  explain the poor convergence of our Monte Carlo simulations the
  small-$\lambda$ regime.
  This data is for system of size $L=8$, $C_0=1000$, ($\rho_0$, $\rho_L$)= (0.6, 0.4)
}
\end{figure}

The near-zero eigenspectrum of the transition rate operator, and its
dependence upon $\lambda$ for fixed boundary conditions, is extremely
interesting in its own right, as shown in Fig.~\ref{fig:eigenspec}.
The structure is rich, with eigenvalues frequently crossing over each
other, as well the appearance and disappearance of bands.

Repeat calculations show that large changes in $C_0$ cause very small
changes in the low-lying eigenspectrum of the transition rate matrix,
so in this sense the slow dynamics are insensitive to $C_0$. It does
affect the group of eigenvalues of larger magnitude,
which form a u-shape towards the top of Fig.~\ref{fig:eigenspec},
suggesting that this group is associated with rapid changes in
configuration at the boundaries. The
extremely large gap between the low-lying and upper sequences of
eigenvalues reinforces that notion.

One of the most notable features of the eigenspectrum is the presence
of a relatively small number of low-lying eigenvalues which seem to
split off from the main sequence as $\lambda$ becomes much smaller
than $1$; they have $\mathcal{O}(\lambda)$ scaling, whilst the main
sequence has $\mathcal{O}(1)$ scaling. This split coincides with, and
may be related to, a suspected transition, which we will discuss in
more detail in Section~\ref{sec:transition}.  System size variations
suggest that the number of eigenvalues which belong to this grouping
scale as $\mathcal{O}(2^{\sim 0.8L})$, slower than the total number of
eigenvalues, $2^{L+4}$, suggesting that they describe long-range
phenomena.  The split begins as the lowest non-zero eigenvalue takes
its lowest values: this may correspond to a change in the character of
the flow (i.e. eigenvector of the zero-eigenvector) to include more
microstates corresponding to high particle density.


Unfortunately the space complexity of the sparse transition rate
matrix scales as $\mathcal{O}(2^L L)$, whilst the matrix density
scales as $\mathcal{O}(2^{-L} L)$.  This makes computations rapidly
become unfeasible for large $L$. The overall time complexity for this
process is $\mathcal{O}(2^{2L}L)$.  In spite of this, we have managed
to compute properties such as the steady-state flow rate for systems
with sizes as large as $10$ internal lattice sites, which are large
enough to produce consistent outputs comparable to our other methods
(as shown in Figures~\ref{fig:lambdaScans} and \ref{fig:wideScans}),
whilst being computationally cheap enough to allow us to perform
calculations with many different parameter configurations.

\section{Simulations}
\label{sec:simulations}
Given access to large-scale computational hardware, it is relatively straightforward to perform Monte Carlo simulations of the SPM, with a little adjustment to the precise implementation of the model. 
We chose to calculate primarily by using the \texttt{KMCLib}~\cite{leetmaa2014kmclib}
package, which implements the Kinetic Monte Carlo algorithm
(essentially the same as the Gillespie algorithm~\cite{Gillespie1977,
 Bortz1975, Prados1997}) on lattice systems. The codes used are kept
here~\cite{jHellGitRepo}.

In the bulk, the transition rates are simply
those described in Fig.~\ref{fig:rates}. At the boundaries we used a similar two-layer boundary trick to that described in Sec.\ref{sec:transMatrix}, only this time the particles appear and disappear with rates
\begin{equation}
\sqrt{\lambda\frac{\rho_0}{1-\rho_0}},
\end{equation}
and
\begin{equation}
\sqrt{\lambda\frac{1-\rho_0}{\rho_0}},
\end{equation}
and likewise for $\rho_L$, respectively. Whilst it would be nice for these rates to be large in order to force the boundaries to maintain the correct density more accurately,
we choose to simply keep them proportional to the geometric mean of $1$ and $\lambda$,
otherwise they tend to either happen too rarely or far too frequently, causing the KMC algorithm to be inefficient.

Independent of the KMCLib code, we wrote a simple Metropolis-Hastings
algorithm~\cite{metHastAlg} which randomly selects single particle hops; this is more efficient than using KMC for values of
$\lambda$ which are relatively close to $1$ as the algorithm is simple and the acceptance rate is relatively high, so we can generate better statistics using this method in that regime.
We calculate the flow from the number of particles entering and
leaving the system at the boundaries.  Since the model is defined in
terms of {\it rates}, the flow in ``particles per unit time'' is a well
defined quantity.

\begin{figure}[h!]
\vspace{0em}
\begin{center}
 \begin{tabular}{c}
    \includegraphics[width=1\linewidth]{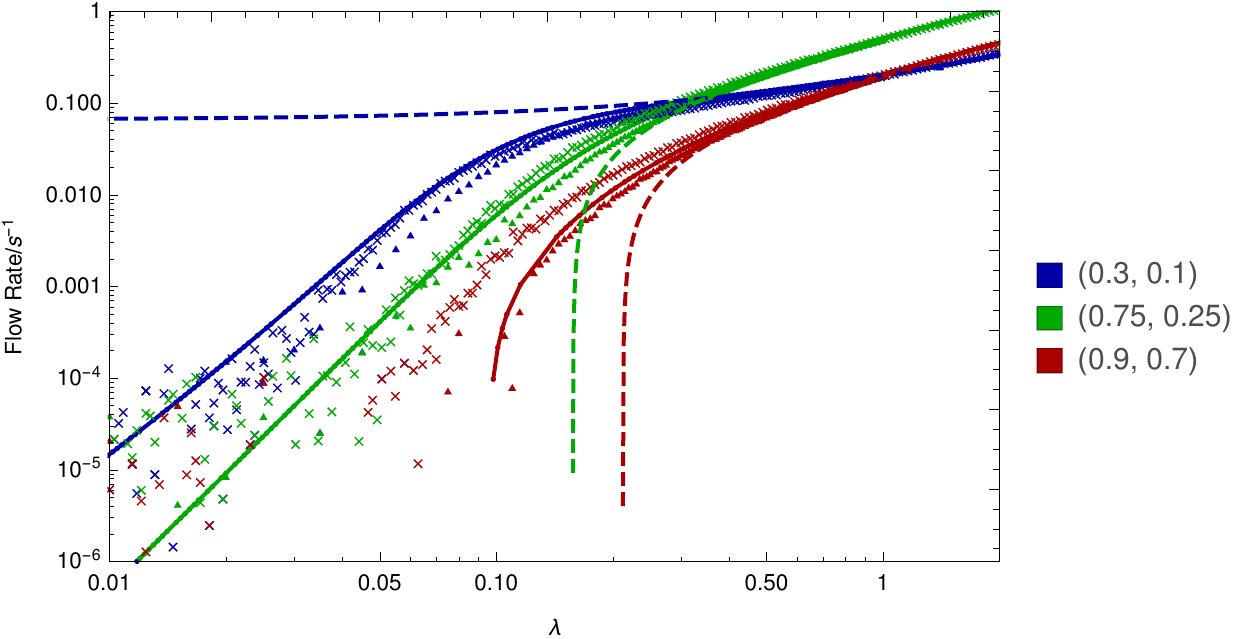} \\
    \includegraphics[width=1\linewidth]{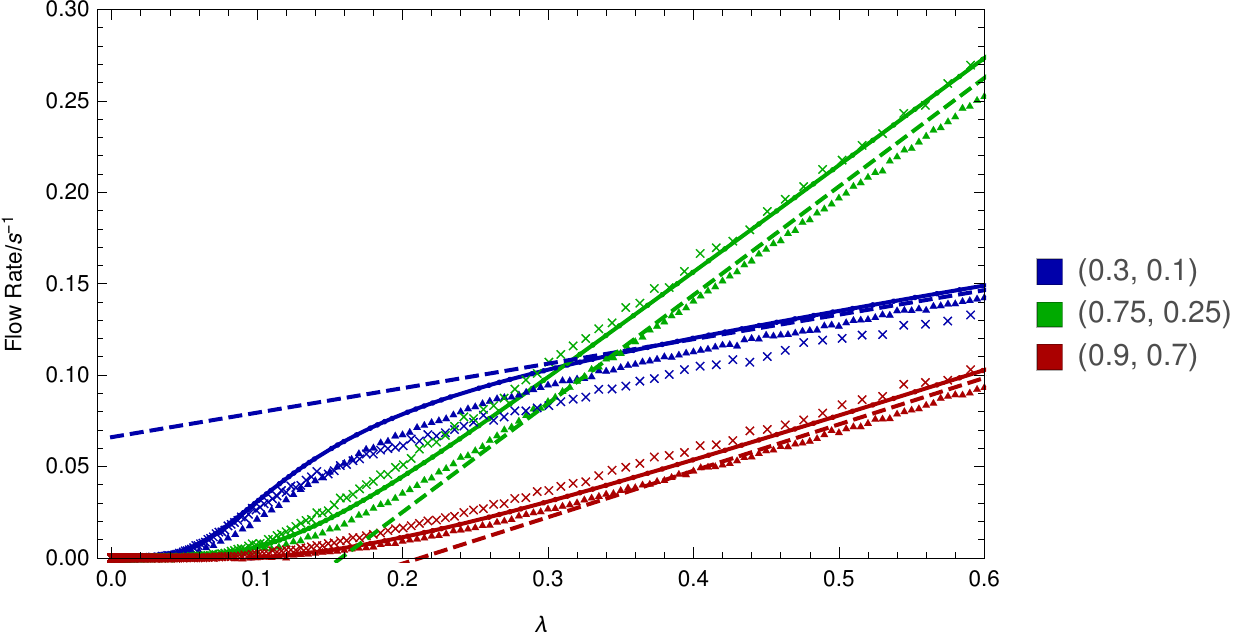}
    \end{tabular}
\end{center}
\caption{\label{fig:lambdaScans} The variation of (normalized) flow rate with $\lambda$ in the SPM with fixed boundary densities, on both logarithmic and linear scales.
Crosses indicate results from the KMC simulations, triangles those from the traditional
Metropolis-Hastings simulations and the joined circles the current observed the TRM method. The dashed lines are the corresponding MFT predictions for the current,
and color indicates the boundary conditions $(\rho_0, \rho_L)$ used. Flow rates are normalized to the driving force, so that we can easily compare different system sizes. The standard error in the flow rate is evident from  the spread of the points on the graph. 
}
    \vspace{0em}
\end{figure}

\begin{figure}[h!]
\vspace{0em}
\begin{center}
    \includegraphics[width=1\linewidth]{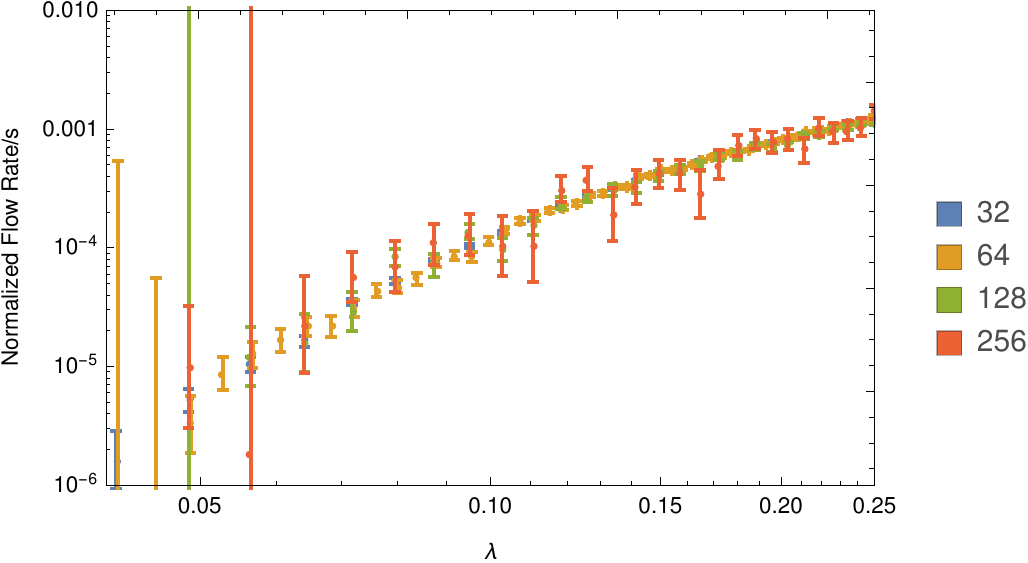}
\end{center}
\caption{\label{fig:sysScaling} The system size dependence of the normalized  flow rate as a function of $\lambda$, computed using the KMC method. These calculations all have boundary conditions $(0.75, 0.25)$, and the flow rate has been
normalized via multiplication by the system size (displayed on legend). Size
$64$ systems are a good compromise in terms of size and accuracy for a given amount of computational time.
}
\end{figure}

Using KMClib we studied systems of length $64$ (lengths $32$, $128$ and $256$ give similar
results as shown in Fig.~\ref{fig:sysScaling}.), running them for $400000$ Gillespie steps for equilibration
followed by $10000$ measurement runs of $1000$ steps interspersed with
relaxation runs of $16000$ steps. This way we could gather statistics
about flow rates and densities in a well-equilibrated
system. Specifically, we generate a pool of $10000$ samples of flow
rate and density, from which we can calculate estimates of the
descriptive statistics of both quantities.
These calculations, in which we held the boundary densities constant whilst varying $\lambda$, were repeated using Metropolis-Hastings for a length $100$ system, and we have transition rate matrix currents
for a system of length $10$. These are plotted on both logarithmic and linear axes in Fig.~\ref{fig:lambdaScans}.
respectively.

\begin{figure}[h!]
\vspace{0em}
\begin{center}
    \includegraphics[width=1\linewidth]{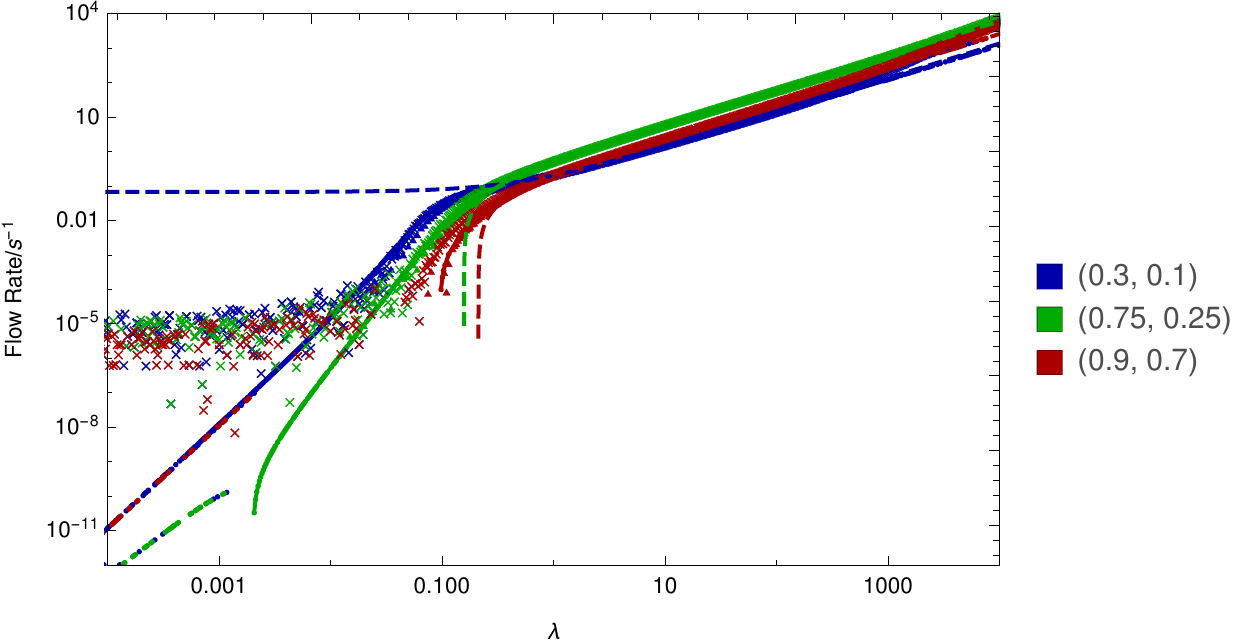}
\end{center}
\caption{\label{fig:wideScans} As the logarithmic plot in Fig.~\ref{fig:lambdaScans}, but over a much wider range of $\lambda$ values. Notice how the data series from the two Monte-Carlo
methods broaden and flatten as $\lambda$ becomes very small; this is because of poor statistics due to the discrete nature of the particles and the extremely low flow rate.
}
    \vspace{0em}
\end{figure}

\subsection{Diffusion Coefficient}
We compare an MFT prediction (see~\ref{sec:mftPred}) and the KMC numerical results for the
diffusion constant in Fig.~\ref{fig:diffCoef}. We see that MFT and
simulation agree well for low stickiness, and both show the symmetry
about $\rho_M = \frac{2}{3}$. For high stickiness, where the MFT
prediction gives a negative diffusion constant, we actually see low
positive values for the current.  It should be noted that the MFT
assumes that $\left\langle \rho \right\rangle = \rho_M$ throughout, whereas in the simulations $\left\langle \rho \right\rangle$ tends to be much higher.

\subsection{Particle Motion During Flow}
It is instructive to get an overview of how the particles move during
flow. Fig.~\ref{fig:flowPatterns} show a plot of the flow structure in
the slow-flow regime.  In very short time averages the ``striped''
pattern indicates separation into dense and sparse regions, with an
overall concentration gradient arising from the relative size of such
regions.  Particle/vacancy diffusion through the empty/full regions
can be seen.  As the averages are taken over longer times the blocks themselves appear to diffuse.  Averaged over the entire simulation (not shown) the averaging simply gives a smooth density gradient.

\begin{figure}[h!]
\begin{center}
 \begin{tabular}{c | c}
    \includegraphics[width=0.49\linewidth]{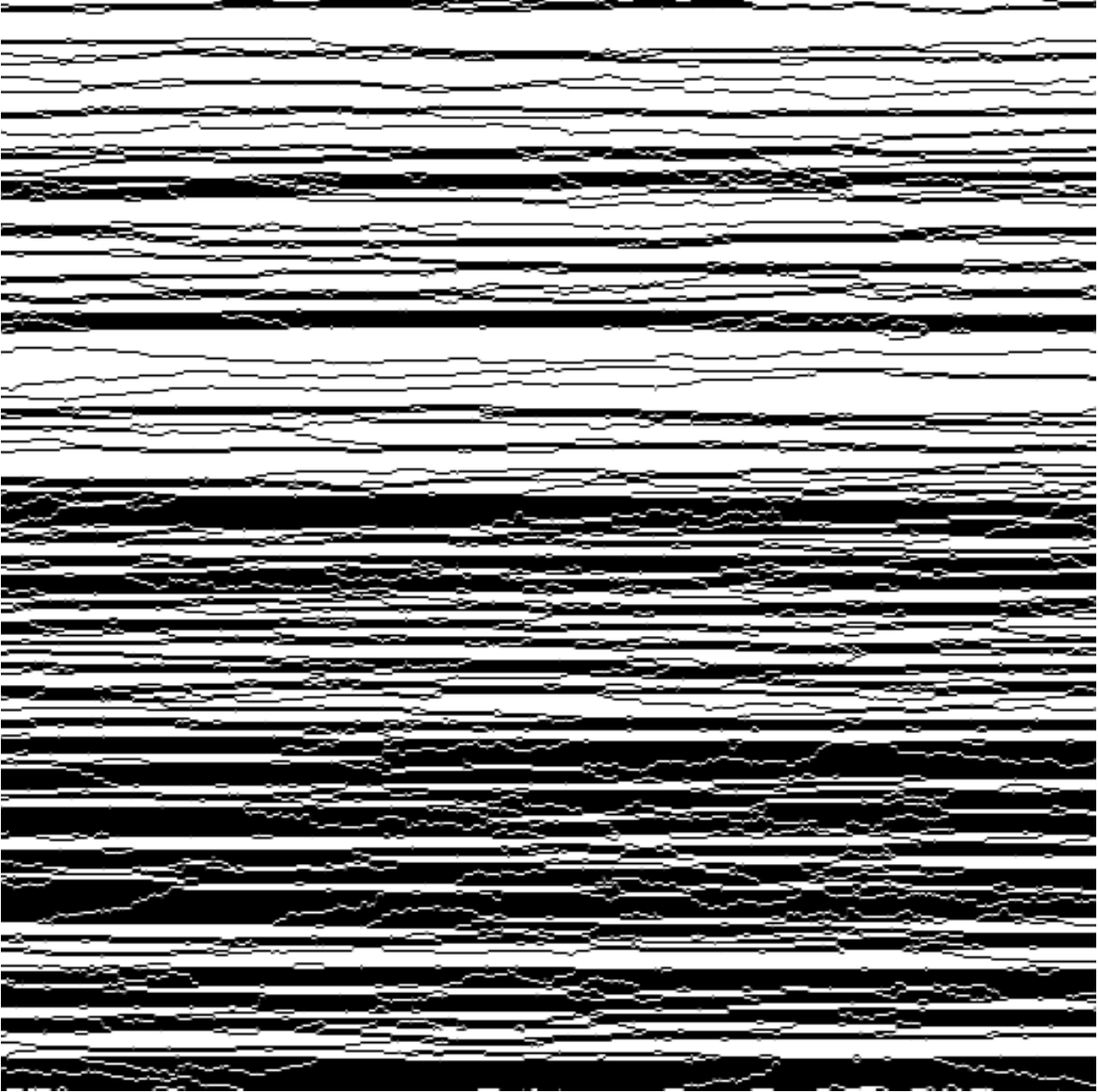}  &\includegraphics[width=0.49\linewidth]{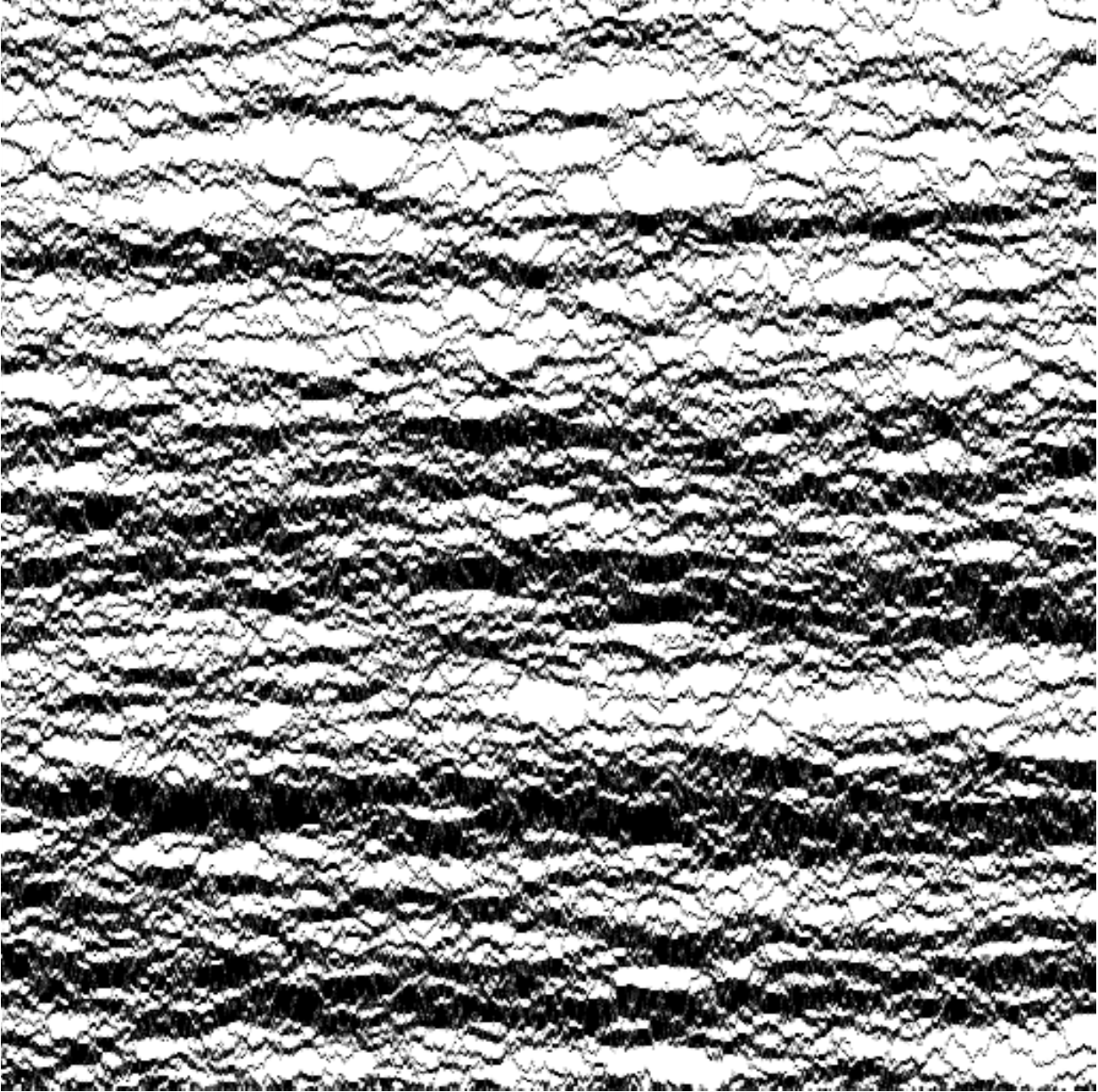} \\
    \hline
    \includegraphics[width=0.49\linewidth]{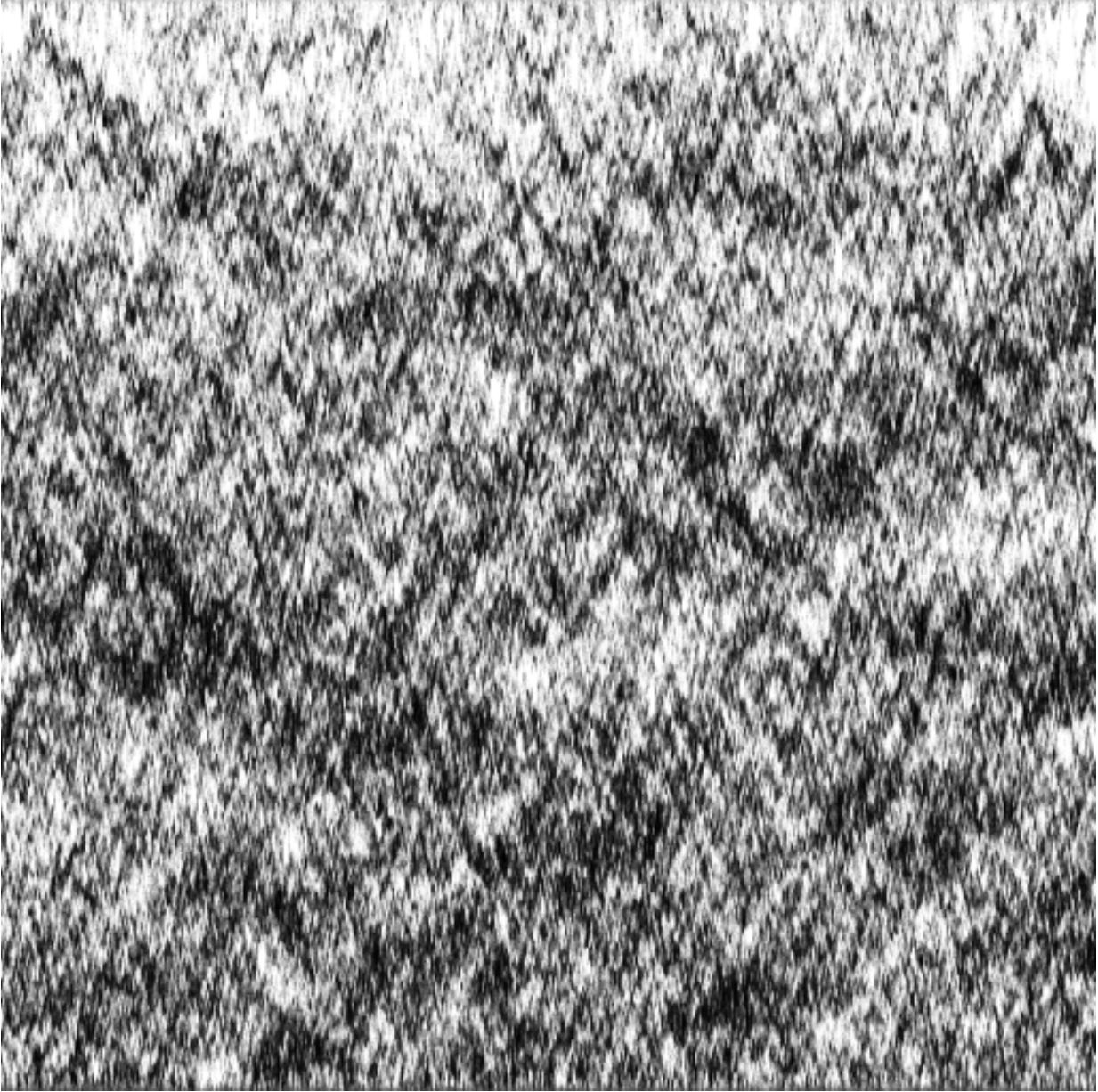} &\includegraphics[width=0.49\linewidth]{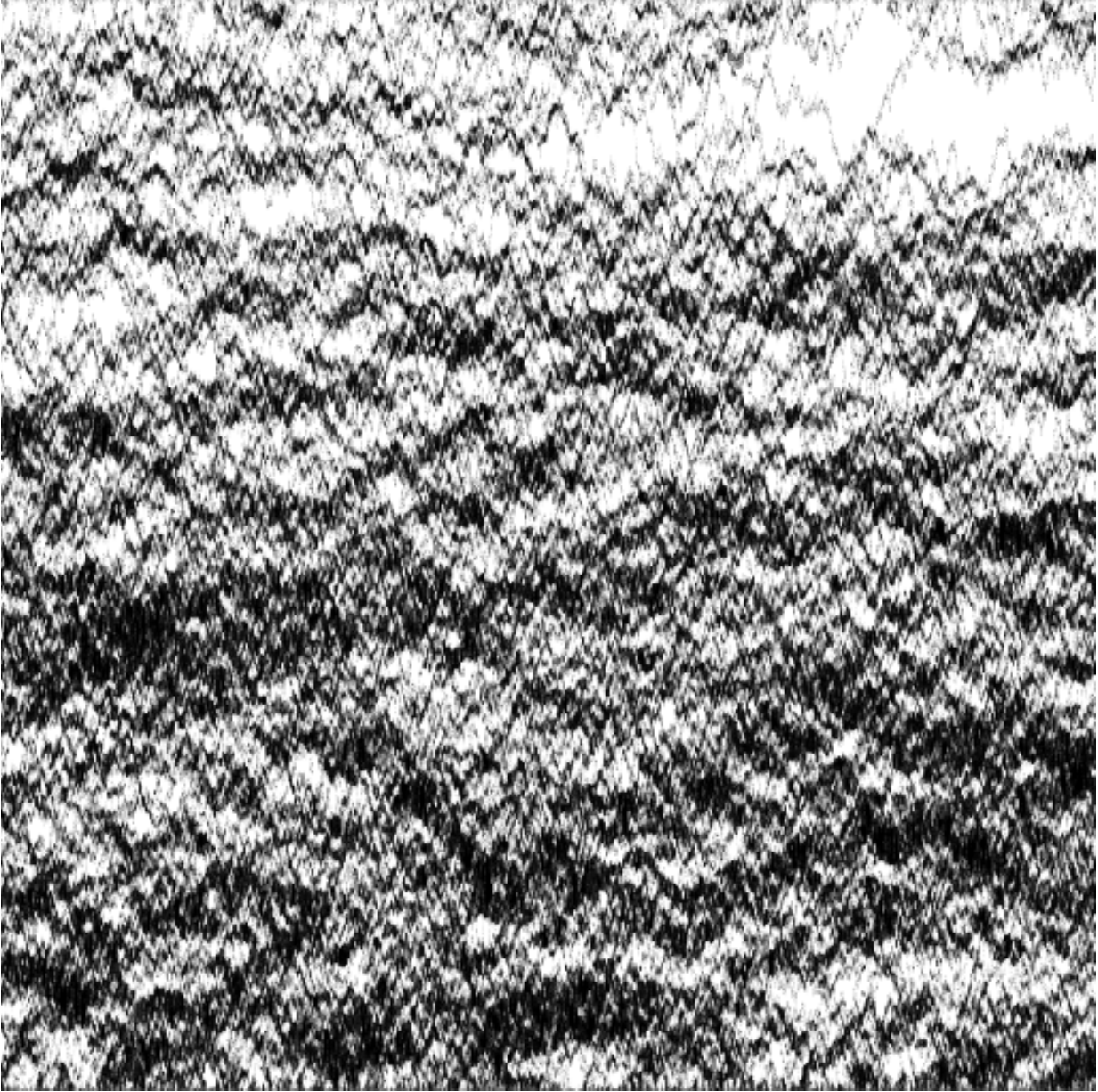}
    \end{tabular}
\end{center}
\caption{\label{fig:flowPatterns} 
Indicative spacetime flow pattern for sticky free-flow $\left[\lambda = \frac{3}{20}, (\rho_0, \rho_L) = (\frac{3}{4}, \frac{1}{4})\right]$; other combinations shown in the supplementary materials.
Time runs along the x-axis, space (1 pixel=1 site) along the y-axis, with grayscale tone (black being empty, white being full) illustrating average site occupation over (clockwise from top left) $\frac{1}{32}$, $1$, $8$ and $32$ Gillespie steps per site respectively.}
\end{figure}

Interesting structure is visible. The dynamics look like a random walk with
some tendency for particles to clump; over longer timescales the
diffusive behavior is more evident, with a textured structure
suggesting characteristic velocity of particles or vacancies through
emergent correlated clumps.  Additional plots can be found in the
supplementary materials.

\subsection{Transition in flow character}
\label{sec:transition}
Fig.~\ref{fig:wideScans} shows how the current varies with stickiness
at fixed driving for a very broad range of $\lambda$.  At high $\lambda$ the rate is simply proportional
to $\lambda$ for any forcing.  This result is far from trivial - it
means that the overall flow is determined by the {\it faster} rate
$(\lambda$), not the slower rate $(1)$.  That the MFT averages over the
two is unsurprising (Eq.~\ref{eq:diffCoef}), but for the
simulation to avoid having a ``rate limiting step'' requires the
system to fill to sufficient density that there are always particles
in contact to repel one another, which suggests that the system is performing some kind of self-organization.

\begin{figure}[h!]
\begin{center}
    \includegraphics[width=1\linewidth]{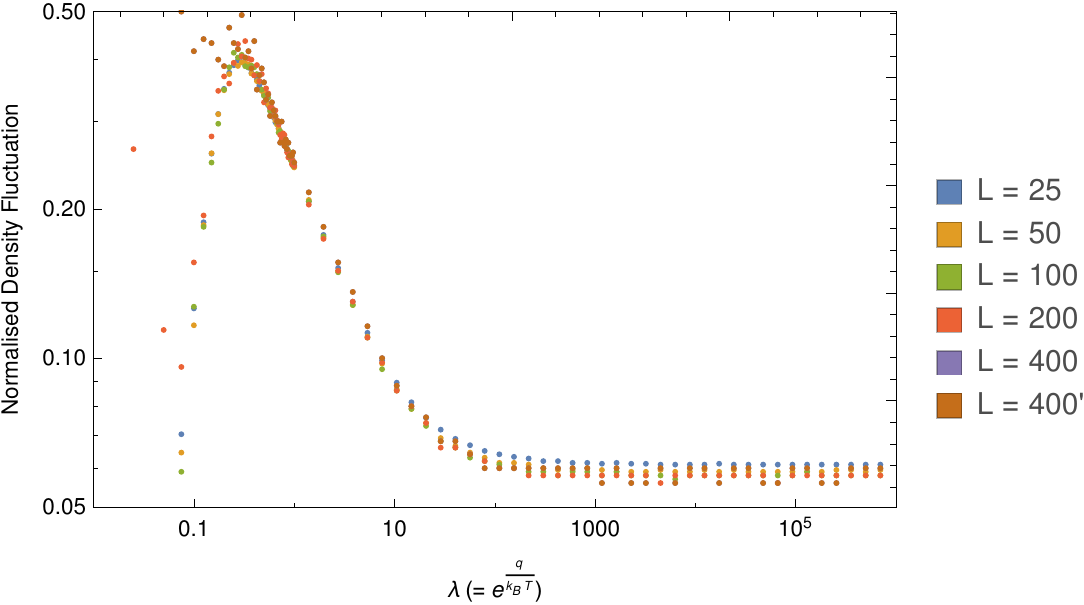}
\end{center}
\caption{\label{fig:DenFluc} The scaling of the normalized fluctuation in the overall system density with system size, as a function of $\lambda$ and with boundary conditions $(0.6, 0.4)$. 
This was computed using the traditional Metropolis-Hastings method. The normalization was achieved by multiplying by the system size, $L$. The great deal of coherence between these results regardless of size suggests
that the fluctuation is of the normal Gaussian type, and not the kind of anomalously-scaling fluctuation associated with equilibrium phase transitions.}
\end{figure}

When $\lambda = \lambda_c\sim0.2$ the simulations show a transition to a different
behavior, with higher or lower $\lambda_c$ depending on the boundary conditions and the strength of the driving force.
The current and its fluctuation remains finite, but there a distinctive peak in the particle density fluctuation
$\left \langle (\rho-\bar{\rho})^2 \right \rangle$ (Fig.~\ref{fig:DenFluc}).  The width of the peak is independent of the
system size 
which suggests a continuous
phase transition from the free flowing to the ``stuck'' regime.

Figs.~\ref{fig:lambdaScans} and \ref{fig:wideScans} reveal that in the $(0.3, 0.1)$ and $(0.75, 0.25)$ boundary
configurations the current switches from being $\mathcal{O}(\lambda)$ at high $\lambda$ to being
$\mathcal{O}(\lambda^k)$, $k\sim4$ for low $\lambda$. In the high-density case, with boundary densities $(0.9, 0.7)$,
the current actually starts flowing backwards for low $\lambda$ in the transition rate matrix calculation. Therefore, we propose that the
peak in the density fluctuation and the odd behavior in the current suggest the existence of a nonequilibrium
phase transition in the SPM.

\subsection{Effective diffusion constant}

\begin{figure}[h!]
\vspace{0em}
\begin{center}
 \begin{tabular}{c}
    \includegraphics[width=0.7\linewidth]{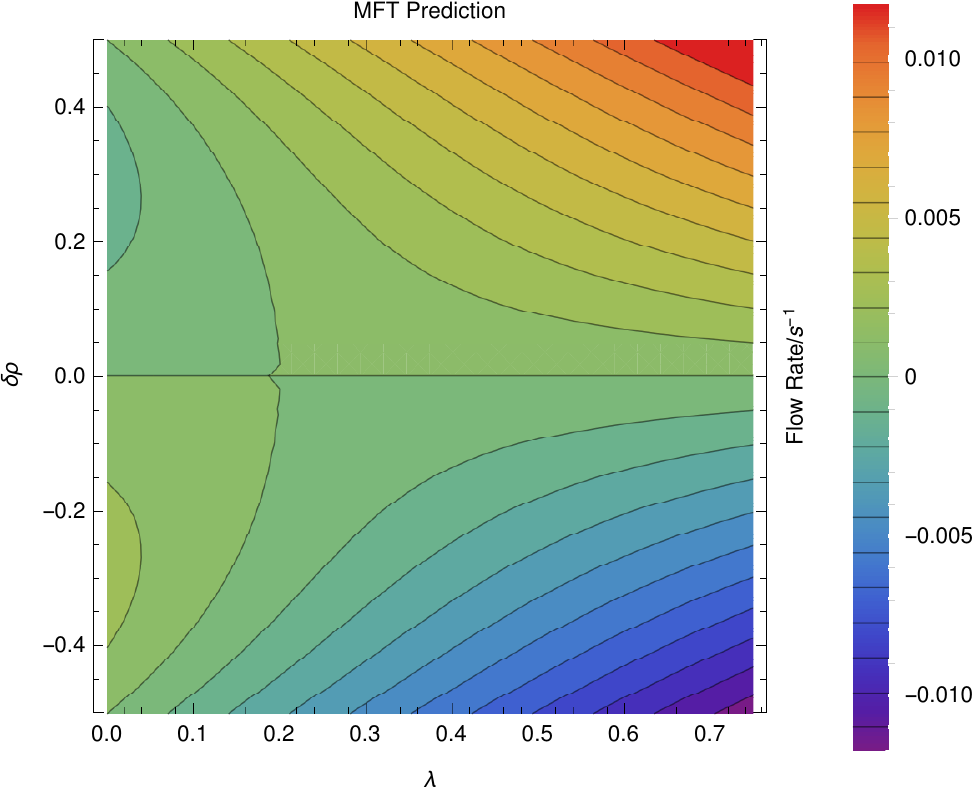} \\
    \includegraphics[width=0.7\linewidth]{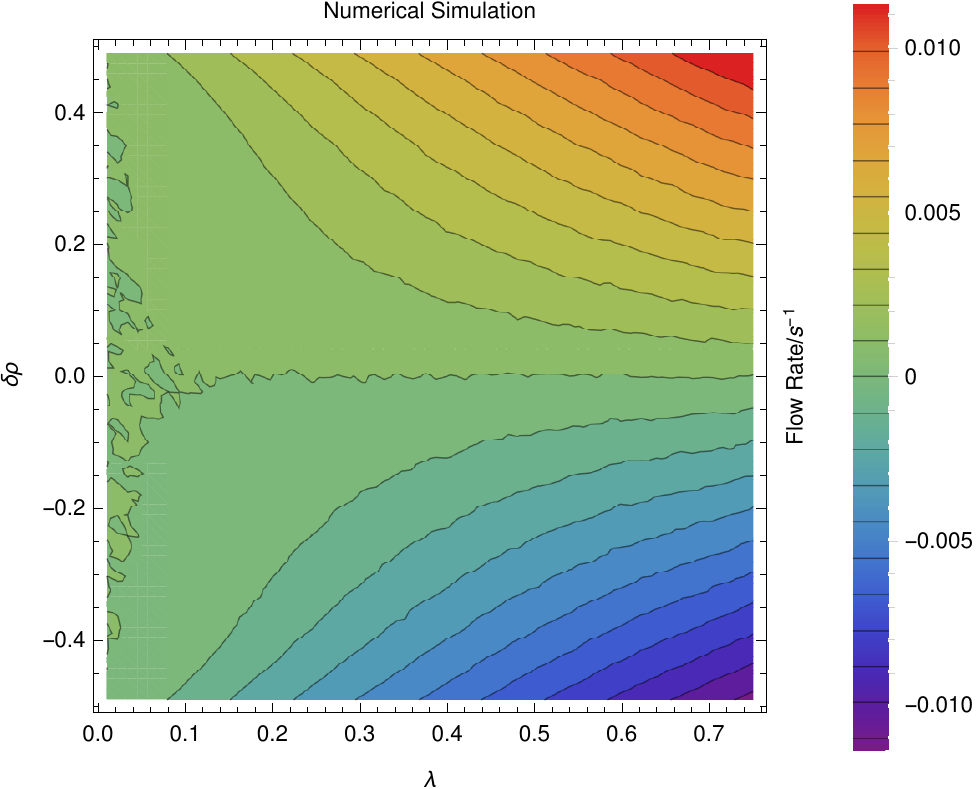}
    \end{tabular}
\end{center}
    \vspace{0em}
\caption{\label{fig:constDens} Flow rate mean observed when varying the difference $\delta\rho$ between the boundary concentrations
$(\rho_0, \rho_L) = (\rho_M + \frac{1}{2} \delta\rho, \rho_M - \frac{1}{2} \delta\rho)$ and $\lambda$ (The top panel is the MFT prediction
for the flow rate, whilst bottom shows the observed mean flow rate).
We chose $\rho_M=\frac{1}{2}$, as this gives us the biggest range of $\delta\rho$ to investigate.
These calculations were performed with the same run parameters (system length etc) as described in Fig.~\ref{fig:diffCoef}.
The MFT prediction (top) shows a region of
negative flow - this occurs at $\zeta > \frac{4}{5}$ ($\lambda < 0.2$) for the weakest
driving force, with additional stickiness required when the driving
force increases. The simulation exhibits close to zero flow in that
region (bottom).  Away from this region, the MFT and
the simulation are in good agreement.
}
\end{figure}

\begin{figure}[h!]
\vspace{1em}
\begin{center}
 \begin{tabular}{c}
    \includegraphics[width=0.7\linewidth]{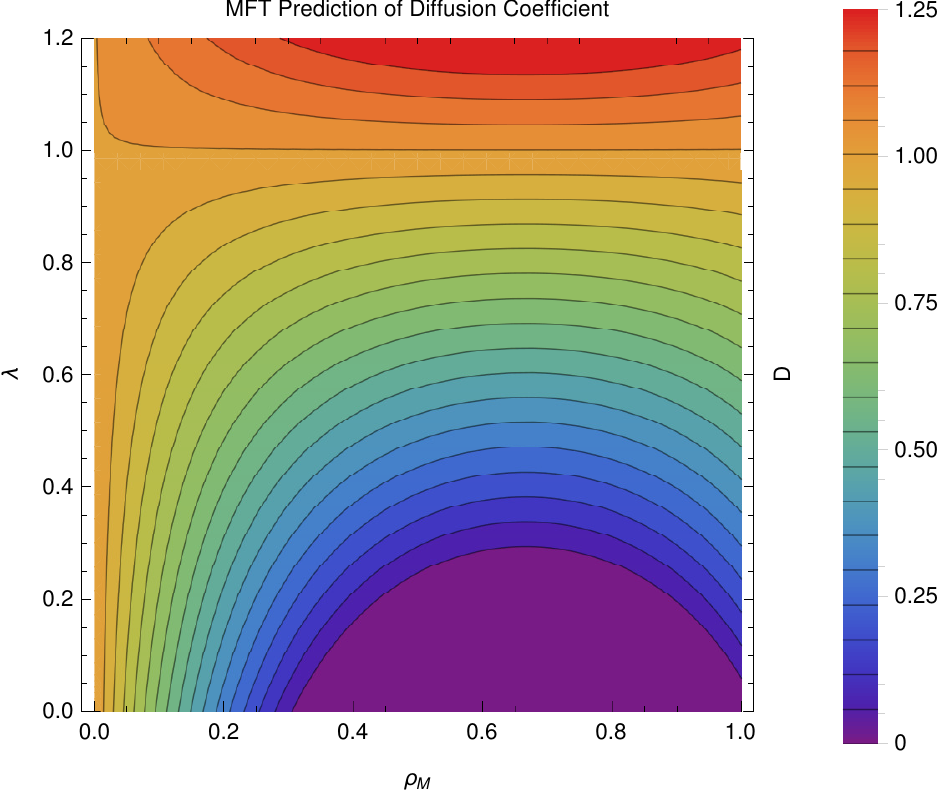} \\
    \includegraphics[width=0.7\linewidth]{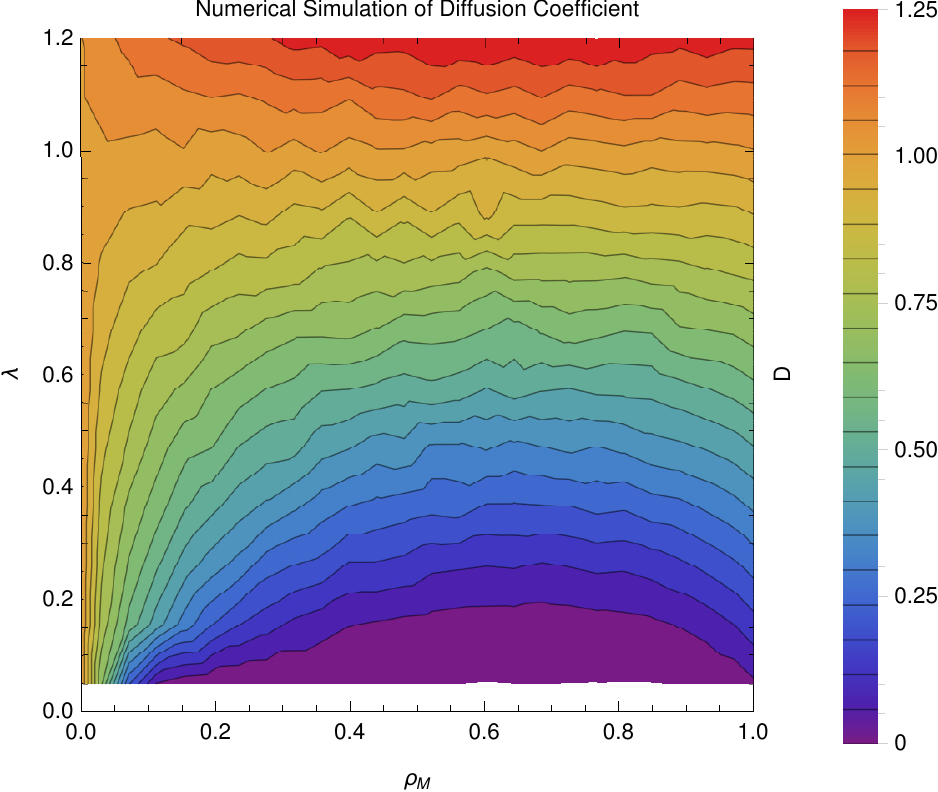}
    \end{tabular}
\end{center}
\caption{\label{fig:diffCoef}
Comparison of effective diffusion coefficient $D$ in the MFT (top) and in direct simulation (bottom) as a function of density and stickiness.
The white region is where the MFT gives negative diffusion. The simulations used 124 sites averaged over $\sim 10^9$ steps at each of $12 \times 24 \times 16 $ $(\lambda, \rho_M, \delta \rho)$ combinations.  
Full details in the supplementary materials.}
    \vspace{-2em}
\end{figure}

Using boundary conditions
\begin{equation}
(\rho_0,\rho_L)=(\rho_M+\frac{1}{2}\delta\rho, \rho_M-\frac{1}{2}\delta\rho) 
\end{equation}
we demonstrate the
dependence of current on boundary density difference and $\lambda$ (Fig.~\ref{fig:constDens}).  This
shows that the transition to the stuck phase is suppressed by stronger
driving forces (large $|\delta\rho|$).
We use the limit of small $\delta\rho$ to calculate the effective
diffusion constant (Fig.~\ref{fig:diffCoef}).  This is normalized to 1 in the case of
$\lambda=1$, which is just SEP.  The $\rho_M=0$ limit corresponds to
free flow of particles, so the diffusion constant here does to
1. Similarly the $\rho_M=1$ limit is flow of vacancies, so
$D\rightarrow\lambda$.  One might expect a monotonic variation between
these limits, but surprisingly the simulations show that there is
always an extremal value for $D$ close to $\rho_M=2/3$: this is a
minimum for $\lambda<1$ and a maximum for $\lambda>1$.

\subsection{Self-organized density}

We calculated the time-averaged total number of particles in
the system by updating a histogram of particle numbers
as the simulation progresses. In each of our calculations, we make the
initial configuration by randomly filling the system with particles
and vacancies in such a way that the initial density should be
$\frac{1}{2}(\rho_0 + \rho_L)$, and then run the system for a
sufficient number of equilibration steps to destroy any initial
transients.

\begin{figure}[h!]
\vspace{0em}
\begin{center}
 \begin{tabular}{c}
    \includegraphics[width=0.9\linewidth]{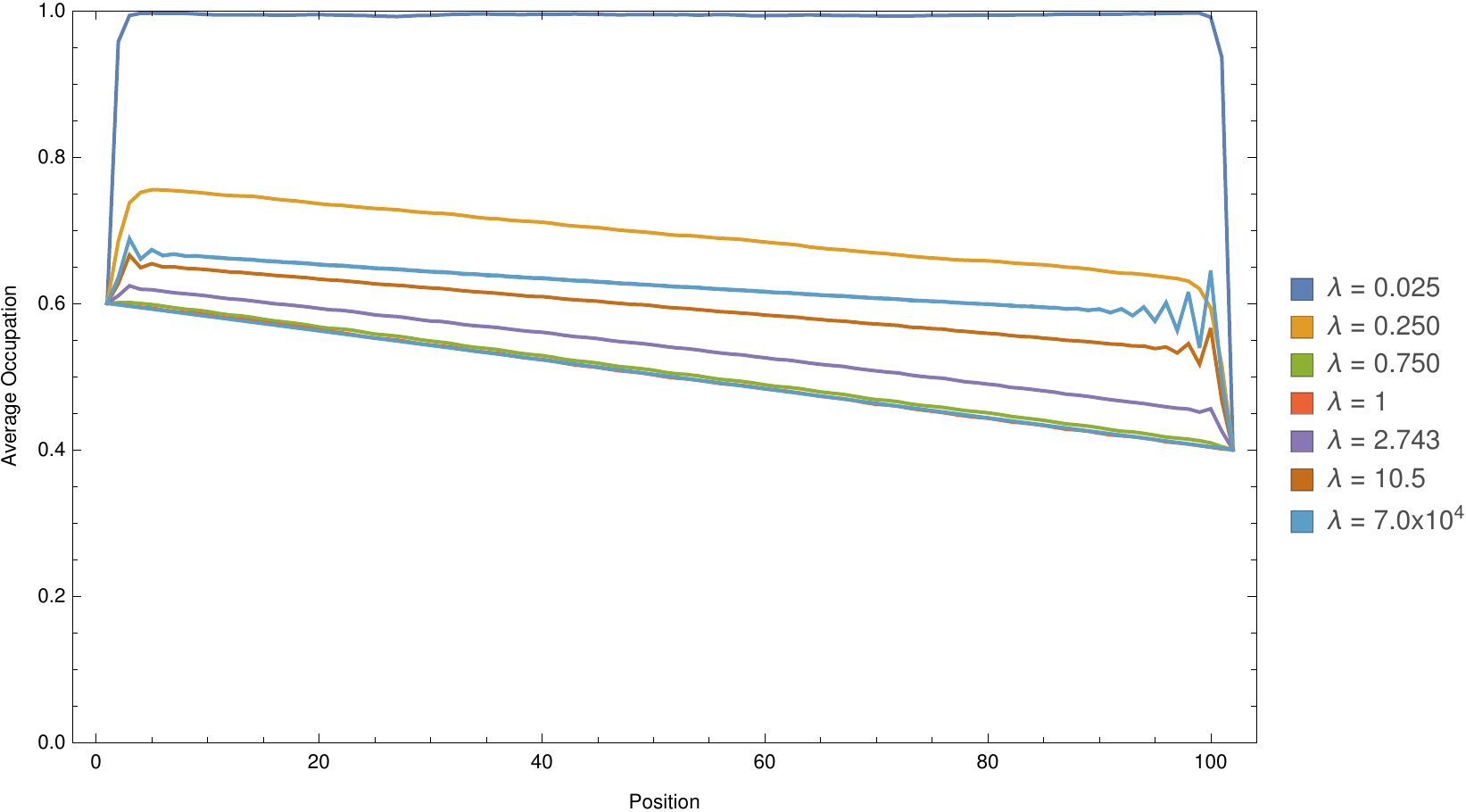} \\
    \includegraphics[width=0.9\linewidth]{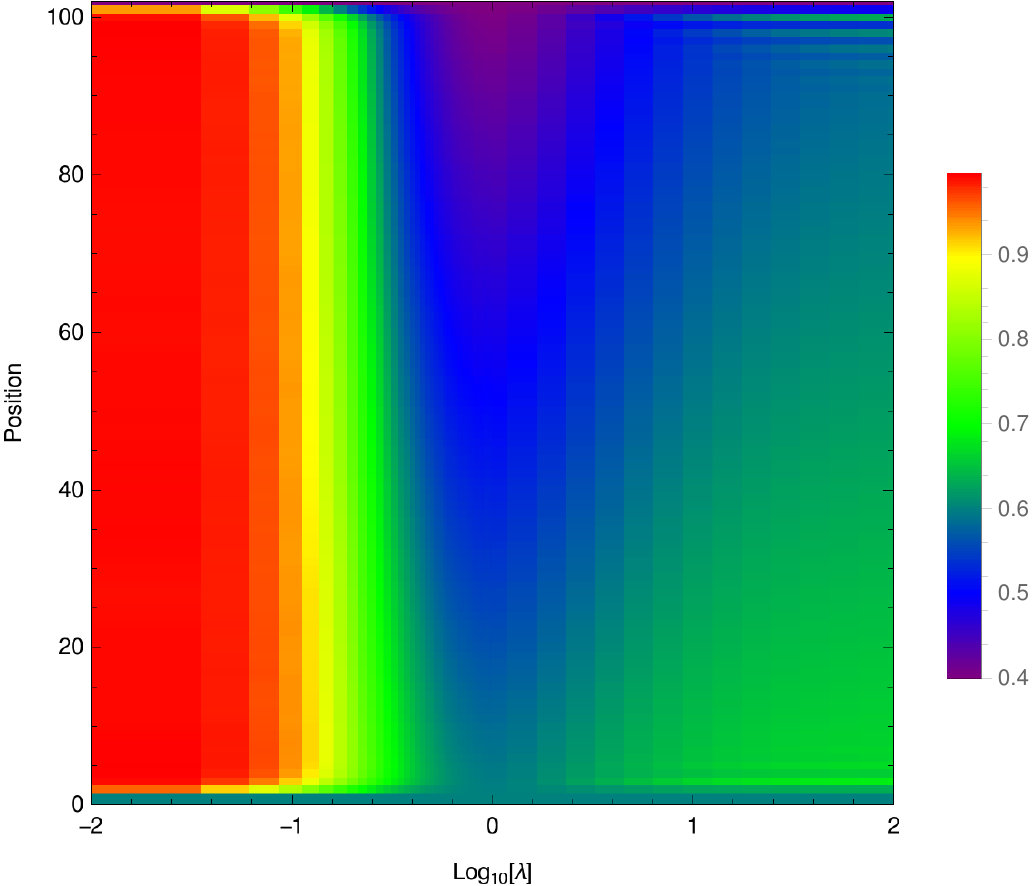}
 \end{tabular}
\end{center}
    \vspace{-0em}
\caption{\label{fig:densProfiles}  The variation of the time-averaged density profile with respect to $\lambda$. The top panel shows this profile for a selection of values for
  $\lambda$ in a system of size $100$, with boundary conditions $(0.6, 0.4)$; the bottom panel displays this variation in the form of a colored density plot.
  Again, we have used the traditional Metropolis-Hastings method for these calculations. The oscillatory behavior near the boundary is a robust feature of the profile at high $\lambda$.}
\end{figure}

\begin{figure}[h!]
\vspace{0em}
\begin{center}
    \includegraphics[width=1\linewidth]{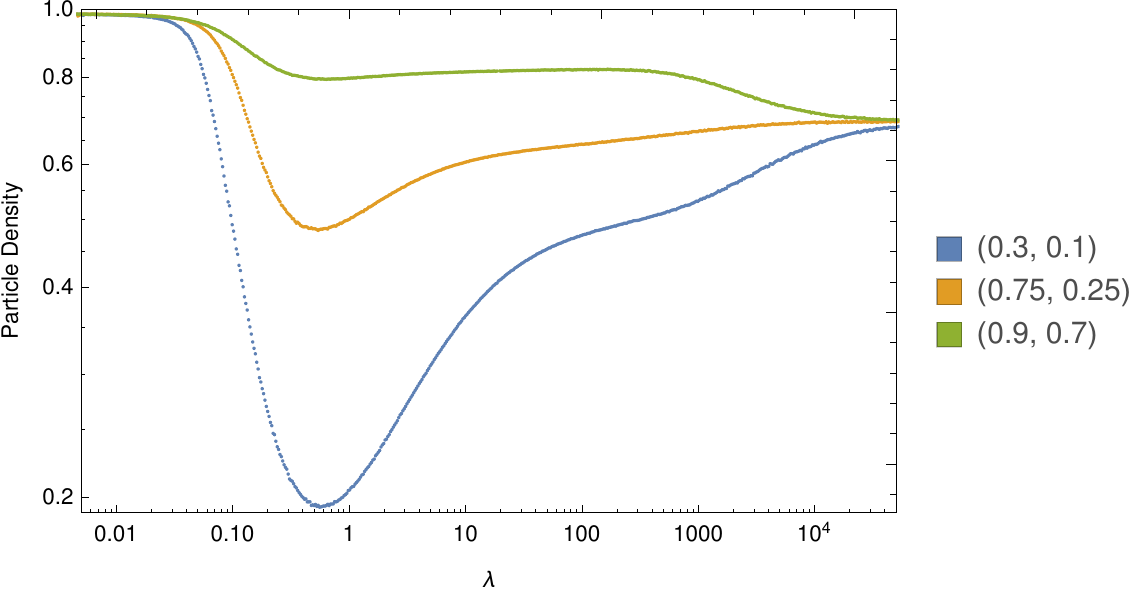}
\end{center}
    \vspace{-0em}
\caption{\label{fig:wideDensities} The overall average density of a system of size $64$, computed using KMC, with boundary conditions as specified in the plot. Notice how the density converges to $1$ and $\sim \frac{2}{3}$
for extreme low and high $\lambda$ respectively.}
\end{figure}

In SEP, ($\lambda=1$) the density varies linearly across the system
from $\rho_0$ to $\rho_L$, as one expects for a diffusion process.
However, for sticky particles, $\lambda<1$, the density rises sharply
near to the boundary, then has a linear profile anchored about a value higher
than $\rho_M$  (Fig.~\ref{fig:densProfiles}). 
One might view this as particles being sucked into the
system to lower their Hamiltonian Energy, but such a notion can be
dismissed since the internal density {\it also rises for repelling
  particles}, $\lambda>1$.  The asymptotic values for the bulk density at
high and low $\lambda$ appear to be 1 and $\frac{2}{3}$ respectively, regardless of the 
mean boundary density $\rho_M$ (see Fig.~\ref{fig:wideDensities}). 

\subsection{Review of the simulations}

MC simulation and TRM analysis of the sticky particle model reveals a number of curious
and unexpected features which have no equivalent in either SEP or the
1D Ising model.

\begin{itemize} \item 
A transition in the dependence of the current on $\lambda$ from $\propto \lambda$ to $\propto \lambda^4$.  
\item
The mean density of the system  has complicated variation with  $\lambda$.  
\item 
The effective diffusion constant has an extremal value at
  intermediate boundary density.
\end{itemize}

To obtain further understanding of these behaviors, we tackle the system
analytically, using mean field theory. Although MFT isn't a particularly
sophisticated technique, it does give us a benchmark which we can compare to our other methods.

\section{Mean Field Theory for Flow} \label{sec:mftPred}
  We will be working in the style described in Sec.~2.2 of \cite{blytheEvans2007} and Sec.~5.2.2 of \cite{AppertRolland2015}.
Let the spacing between lattice
sites be $a$, let $\tau_0$ be the free-particle hopping timescale, and
the time-averaged (or ensemble-averaged, assuming ergodicity)
occupation probability of the $i^{\mathrm{th}}$ lattice site be
$\rho_i$.  We will introduce $\zeta = 1 - \lambda $ here for convenience.

\subsection{Lattice Mean Field Theory} Let the ensemble-averaged occupation
probability of the $i^\mathrm{th}$ site at time $t$ be $\rho_i
(t)$. In the mean-field approximation this is assumed to be
independent of $\rho_j(t)$ for $j \neq i $ at equal times. Therefore,
if the $i^\mathrm{th}$ site is occupied, then the rate at which it empties
is the sum of contributions from the
four permutations of the $(i+1)^\mathrm{th}$ and
$(i-1)^\mathrm{th}$ occupancy: 

\begin{align}
\begin{split}
 &\frac{1}{\tau_0 } (1-\rho_{i-1})\left[ (1 - \rho_{i+1}) + \lambda \rho_{i+1} \right] \\
 +&\frac{1}{\tau_0 } (1-\rho_{i+1})\left[ (1 - \rho_{i-1}) + \lambda \rho_{i-1} \right] .
\end{split}
 \end{align}

Similarly, if the $(i)^\mathrm{th}$ site is unoccupied, it fills with
rate which depends on the occupation probability of the neighbours,
and whether they are stuck to the $(i+2)^\mathrm{th}$ and $(i-2)^\mathrm{th}$ sites
respectively. Therefore, an unoccupied $i^\mathrm{th}$ site fills with rate
\begin{equation}
\frac{1}{\tau_0 } \left\{ \rho_{i+1} \left[ \lambda \rho_{i+2} + (1-\rho_{i+2}) \right] + \rho_{i-1} \left[ \lambda \rho_{i-2} + (1-\rho_{i-2}) \right] \right\}.
\end{equation}

If we now multiply the filling/emptying rates of site $i$ by the
probability of being empty/full respectively, we obtain the
final equation for the site occupation 
\begin{align}
\label{eq:latticeMFT}
\begin{split}
 \tau_0 \partDeriv{\rho_i}{t} &= \left( 1-\rho_i \right) \left[ \left(1-\zeta\rho_{i-2} \right) \rho_{i-1} + \left(1-\zeta\rho_{i+2} \right) \rho_{i+1} \right] \\
 &- \rho_i \left[ 2 \zeta \rho_{i-1} \rho_{i+1}  - (3-\zeta)\left(\rho_{i-1} + \rho_{i+1}\right) + 2 \right].
 \end{split}
 \end{align}

\subsection{Diffusion Equation: MFT Continuum Limit}
 To obtain the continuum limit of the MFT we substitute $\rho_i(t)
 \rightarrow \rho(x, t)$, $\rho_{i+m}(t) \rightarrow \rho(x + am, t)$
 into Eq.~\ref{eq:latticeMFT}.  Then, using a Taylor expansion around $x$ for
 small $a$, neglecting terms of $\mathcal{O}(a^4)$, and collecting
 terms we find that
\begin{align}
 \begin{split}
  \tau_0 \partDeriv{\rho}{t} =& a^2 \left[ 1-\zeta \rho (4-3\rho)  \right] \partDeriv{^2 \rho}{x^2} 
\\
  +& 2 a^2 \zeta (3\rho-2) \left(\partDeriv{\rho}{x}\right)^2 + \mathcal{O}(a^4) ,
 \end{split}
\end{align}
which can be factorized into the more familiar form of a continuity equation
\begin{equation}
\label{eq:contPDE}
 \partDeriv{\rho}{t} = \frac{a^2}{\tau_0} \partDeriv{}{x} \left\{ \left[1 - \zeta \rho\left(4-3\rho\right) \right] \partDeriv{\rho}{x} \right\},
\end{equation}
having dropped the higher-order terms.  From this we can identify the
flow $J(x)$ as the term in the curly brackets, and define a mean field
diffusion constant (see Fig.\ref{fig:diffCoef}):

\begin{equation} D(\zeta,\rho) =  \frac{a^2}{\tau_0} \left[ 1 - \zeta \rho\left(4-3\rho\right) \right ] \label{eq:diffCoef} \end{equation}

We note that the MFT diffusion constant can become negative. 
The density at
which this occurs is given by finding the roots $\rho_c$ of the RHS of Eq.~\ref{eq:diffCoef}:
\begin{equation}
\rho_c=\frac{2}{3} \pm \frac{1}{3} \sqrt(4-3/\zeta). 
\end{equation}
The possible values for $\rho_c$ are only real (and therefore physically possible) for $\zeta<0.75$.  So in the limit
of very sticky particles ($\lambda<0.25$) the MFT predicts a
transition between forward and backwards diffusion at some densities.

\subsection{Limiting Cases}

In order to understand the implications of the MFT, let us consider
some limits. As $\zeta \rightarrow 0$ (i.e. as the model becomes a
simple exclusion model), $D \rightarrow \frac{a^2}{\tau_0}$. Likewise,
in the dilute limit $\rho \rightarrow 0$, $D \rightarrow \frac{
  a^2}{\tau_0}$, reflecting the fact that it becomes a dilute lattice
gas and therefore the interactions between particles become irrelevant
as they never meet.  Conversely, in the full limit $\rho \rightarrow
1$, $D \rightarrow \frac{\lambda a^2}{\tau_0}$; this is because we now
have a dilute gas of vacancies, which hop with rate
$\frac{\lambda}{\tau_0}$.  

One may observe that the 
MFT has a symmetry under $\rho \mapsto \frac{4}{3} - \rho$; thus, the
dynamics should be symmetric under a density profile reflection around
$\rho = \frac{2}{3}$. This is where $D$ always attains its extremal
value, $ \frac{a^2}{\tau_0}\left[1 - \frac{4}{3}\zeta\right]$, hence
for $\zeta>3/4$ the diffusion coefficient becomes negative in regions
with $\frac{2}{3} - \frac{\sqrt{\zeta\left(4\zeta - 3\right)}}{3\zeta}
< \rho < \frac{2}{3} + \frac{\sqrt{\zeta\left(4\zeta -
    3\right)}}{3\zeta}$. 

Finally, it is possible to show that
solutions to the continuum MFT containing domains with a negative
diffusion coefficient are linearly unstable; thus, if we try to have a
flow containing $\rho$ for which $D(\rho)<0$, the density of the
medium should gravitate towards one of the two densities for which $D(\rho)\sim
0$. Instead of observing ``backwards diffusion'' we would see an
extremely slow flow or no flow at all. 

\subsection{Steady State Flow}

It is possible to solve the continuum MFT in a steady state on a
finite domain, say $x\in(0, L)$. Steady state implies that there is no
build-up of particles $\partDeriv{\rho}{t}=0$, so the flow is constant through the system.

Using the constant-flow requirement for continuity
$J(x)=J_0=\mathrm{const.}$, and by integrating both sides of
Eq.~\ref{eq:contPDE} with respect to $x$ we find that
\begin{eqnarray}
 \int_{x_0}^x J(x')dx' &=& -\frac{a^2}{\tau_0} \int \! \! \mathrm{d} \rho \left[1 - \zeta
   \rho\left(4-3\rho\right) \right]\\ (x-x_0)J_0 & = & -\frac{a^2}{\tau_0}\left[ \rho + \zeta (\rho -
 2) \rho^2 \right ].
\label{eq:rho_x}
\end{eqnarray}
The density profile across the system is given by solving for
$\rho(x)$.  Since Eq.\ref{eq:rho_x} is cubic, the solution for density
$\rho(x)$ is non-unique for cases of high $\zeta$.  Thus in the limit of
high stickiness, the MFT is unable to make a unique prediction for the
density.  Furthermore, except in the SEP case $\zeta=0$, the density
will not vary linearly across the system.

\subsection{Dirichlet Boundary Conditions}

The constants $x_0$ and $J_0$ relate to the boundary conditions for the
flow, but they need not correspond to a physically-realizable
situation.  For driven systems it is more convenient to consider fixing
the density at each end. 

If we impose such Dirichlet boundary conditions on this system, say
$\rho(0)=\rho_0$ and $\rho(L)=\rho_L$, we find that
\begin{equation}\label{eq:MFTflow}
 J_0 = \frac{a^2}{L \tau_0} \left[ \rho_0 - \rho_L + \zeta \left( \rho_0\left[\rho_0^2-2\right] - \rho_L\left[\rho_L^2-2\right] \right) \right].
\end{equation}

This equation can be used for direct comparison with the simulations
(Fig.~\ref{fig:constDens} and Fig.~\ref{fig:diffCoef}).  In general, the agreement is
good, except for the region where the MFT predicts negative flow.

\subsection{Interpretation of MFT}

The mean field theory enables us to predict flow behavior of the SPM.
It recovers well-known limiting cases such as SEP ($\lambda=0$),
however at high stickiness ($\lambda<0.25$) it predicts its own
demise, with unphysical negative diffusion constants and by having
multiple solutions for the density at some positions in the system,
meaning that the MFT does not give a unique density
profile.  This breakdown of the MFT corresponds to the transition to
slow flow observed in the simulation.

In some
conditions MFT predicts densities greater than 1.  One might guess that when the
MFT offers two possible values for the density, it will correspond to
a phase separation in the actual system. Furthermore,
the MFT prediction of a maximum diffusion constant at a density of
$\rho=2/3$ suggests that this value of $\rho$ might be favored for
strongly driven flows.

A curious feature of the MFT diffusion (Fig.~\ref{fig:diffCoef}) is
that for fixed stickiness there are {\it two possible densities}
giving the same diffusion constant.  Thus it is possible to have a
steady state flow with phase separation into regions of high and low
density. This echoes the situation seen in short time-averages of the
simulation (Fig.\ref{fig:flowPatterns}) where blocks of high and low
density are evident, as well as the oscillatory behavior near the boundaries at high $\lambda$ (Fig.~\ref{fig:densProfiles}). 

Regarding the phase transition, we should note that we only have an
MFT prediction for the flow rate for $\lambda>\frac{1}{4}$, since
$\rho(x)$ stops being unique when $\lambda$ drops below $\frac{1}{4}$. For higher $\lambda$, the MFT is broadly in good agreement with the
simulations, and this continues as $\lambda$ stretches into the
thousands, where the mean flow varies as $\mathcal{O}(\lambda^1)$.

The MFT prediction for the mean flow continues to be a good fit until $\lambda$ becomes sufficiently small,
when the simulations show no evidence of negative diffusion; rather the flow becomes critically slow for very sticky particles.
The higher moments of the simulated flow (e.g. variance) do not show peaks, indicating that hard transitions are not occurring.
Finally, the  simulated density is very close to the average of the boundary densities until $\lambda$ drops below 1/4, at which point the system fills.

\section{Discussion}

We have analysed the model using three methods: Direct simulation,
Transition rate matrices and mean field theory. Each has advantages
and drawbacks, but all show a nonequilibrium transition between a
free-flowing and a blocked phase.  In direct simulation this appears
as a peak in the energy and density fluctuations and a change from
$~\lambda$ to $~\lambda^4$ dependence in the flow rate.  In MFT it
appears as a negative diffusion constant.  In the TRM it corresponds
to a minimum in the minimum eigenvalue, and a crossover in the
character of the associated eigenvector, as clearly illustrated in
Figure.\ref{fig:flowPatterns}

The MFT suggests a symmetry between vacancy-type and particle-type
flow at density of $\frac{2}{3}$, which is observed in the MC
simulations and TRM calculations.  The density self-organizes to this
value in the limit of strongly repelling particles.

The continuum limit solution to
the MFT is a good predictor of the bulk flow behavior of the SPM for
high $\lambda$.  The negative diffusion constant found in MFT at high
stickiness indicates that the assumption of homogeneous density breaks
down: thus the MFT predicts its own demise at a point which agrees well with
the numerics.

Above a certain level of stickiness, the model exhibits a
nonequilibrium phase transition to a slow-flowing phase.  The required
stickiness for the transition is dependent on the strength of the
driving force - large differences between the boundary densities push
the transition towards higher stickiness.  Mean field analysis, together
with visualization of the flowing system, suggest that the transition
occurs when the density becomes inhomogeneous.

The flow exhibits a foamy pattern with time-and-space
correlations of intermediate size.  
The open boundary condition means the system self-organizes its
density.  At zero interaction ($\lambda=1$) this must be the mean of
the boundary conditions, and it is unsurprising that high stickiness
means the system fills.  Strong repulsion between the particles leads
to a density of $\frac{2}{3}$, which appears to maximize the flow rate
at high $\lambda$; there is no known physical principle which requires this.

\section{Conclusions}

The sticky particle model is the combination of the Ising model with
the symmetric exclusion process.  It represents the simplest possible
flow model for interacting particles in a homogeneous medium, and as
such is a model for many physical systems where particles are driven
by a pressure or concentration difference across the boundaries.

We have shown that it exhibits complicated and discontinuous behaviour
absent in either 1D Ising or SEP model, specifically, a transformation
from fast to slow flow with stickiness. 

A number of questions remain open.  Is there a physical principle
which determines the density (Fig.~\ref{fig:densProfiles})?  Why does
the strongly repelling system produce a density with maximum flow
(Fig.~\ref{fig:wideDensities})?  Can one derive the $~\lambda^4$
dependence of the slow-flow (Fig.~\ref{fig:wideScans})?  These clear,
emergent results from our detailed work strongly suggest that there is a higher
level description of complex flow which could predict
them, but to our knowledge no such unifying theory of nonequilibium
thermodynamics exists.

\section*{Acknowledgements}
We would like to thank EPSRC (student grant 1527137) and Wolfson
Foundation and ERC for providing funding, Mikael Leetmaa for producing
\texttt{KMCLib}, and the \texttt{Eddie3} team here at Edinburgh for
maintaining the hardware used.  We would also like to thank Martin
Evans, Bartek Waclaw and Richard Blythe for some very helpful
discussions.

\appendix
\section{Linear Stability of MFT Solution.} Let $\rho_0 (x)$ be a solution to the steady-state continuum MFT, and let us apply a small perturbation $\delta \rho(x, t)$. Let us also assume that $\rho_0$ is also slowly-varying in $x$,
in the sense that $\delta \rho = o(\rho_0)$ and $ \partDeriv{\rho_0}{x} = o(\delta \rho)$, which should be approximately true for a large domain.
Then substituting $\rho = \rho_0 + \delta \rho$ into the continuum-limit MFT and only the highest-order terms, we find that
\begin{equation}
 \partDeriv{\delta \rho}{t} =  \frac{a^2}{\tau_0} \rho_0 (4 - 3 \zeta \rho_0) \partDeriv{^2 \delta \rho}{x^2} + o(\delta \rho).
\end{equation}
Performing a  Fourier transform  (or rather, a suitable local equivalent) with respect to $x$ to yield $\hat{\delta \rho}(k, t) = \mathcal{F}\left[ \delta \rho(x, t) \right]$, this becomes
\begin{equation}
 \partDeriv{\hat{\delta \rho}}{t} = -k^2 \frac{a^2}{\tau_0}\rho_0 (4 - 3 \zeta \rho_0)\hat{\delta \rho}.
\end{equation}
This means that so long as $\zeta < \frac{3}{4}$, the RHS is always negative, and therefore small perturbations decay exponentially and all is well. However, if $\zeta>\frac{3}{4}$, there may be regions of the solution where
$\rho_0 (4 - 3 \zeta \rho_0)$ becomes negative, causing small perturbations to grow exponentially with an emphasis on high-wavenumber (short-lengthscale) modes, which indicates linear instability.

\section{Additional Flow Data.} 
\begin{figure*}[h!]
\vspace{1em}
\caption{\label{fig:fullLambdaScans} Additional flow rate moments and overall system densities for the sweeps through $\lambda$ with fixed boundaries.}
\begin{center}
 \begin{tabular}{c|c}
    \includegraphics[width=0.5\linewidth]{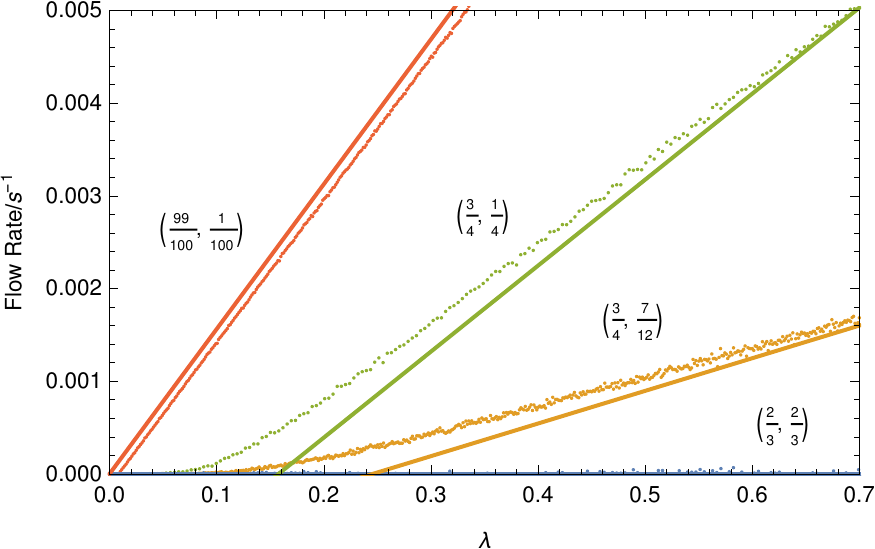} & \includegraphics[width=0.5\linewidth]{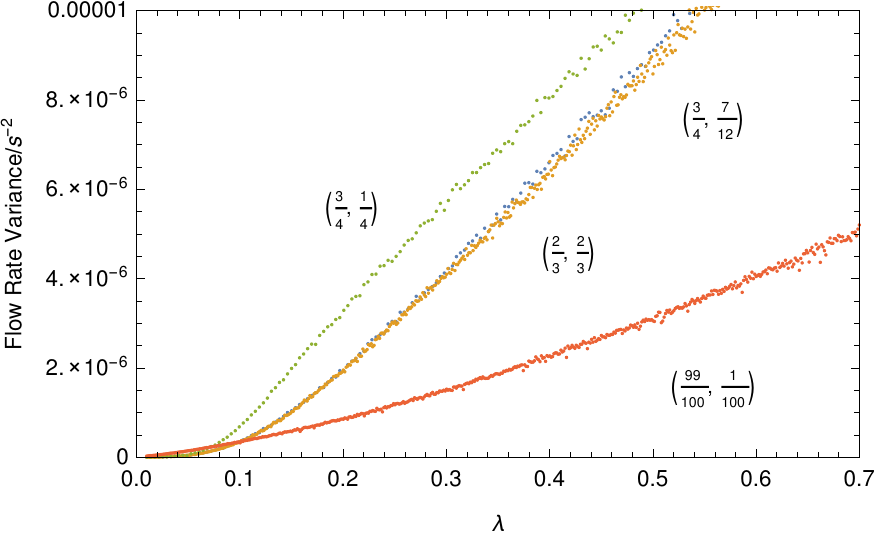} \\
    \hline
    \includegraphics[width=0.5\linewidth]{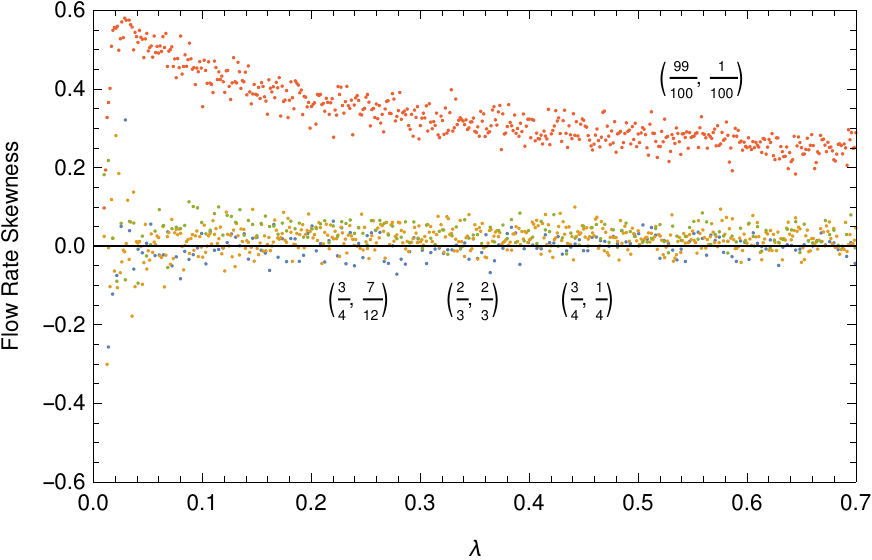} & \includegraphics[width=0.5\linewidth]{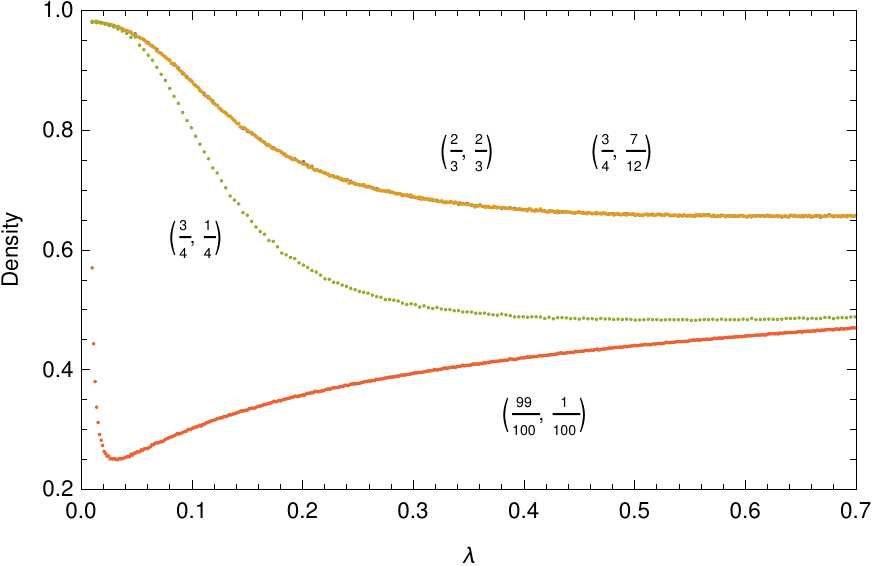} \\
    \end{tabular}
\end{center}
    \vspace{-0em}
\end{figure*}

\begin{figure*}[h!]
\vspace{1em}
\caption{\label{fig:fullConstDens} Flow rate variance and average overall densities observed when varying the difference $\delta\rho$ between the boundary concentrations
$(\rho_B, \rho_T) = (\rho_M + \frac{1}{2} \delta\rho, \rho_M - \frac{1}{2} \delta\rho)$ and $\lambda$. $\rho_M=\frac{1}{2}$, as in the paper.}
\begin{center}
 \begin{tabular}{c|c}
    \includegraphics[width=0.5\linewidth]{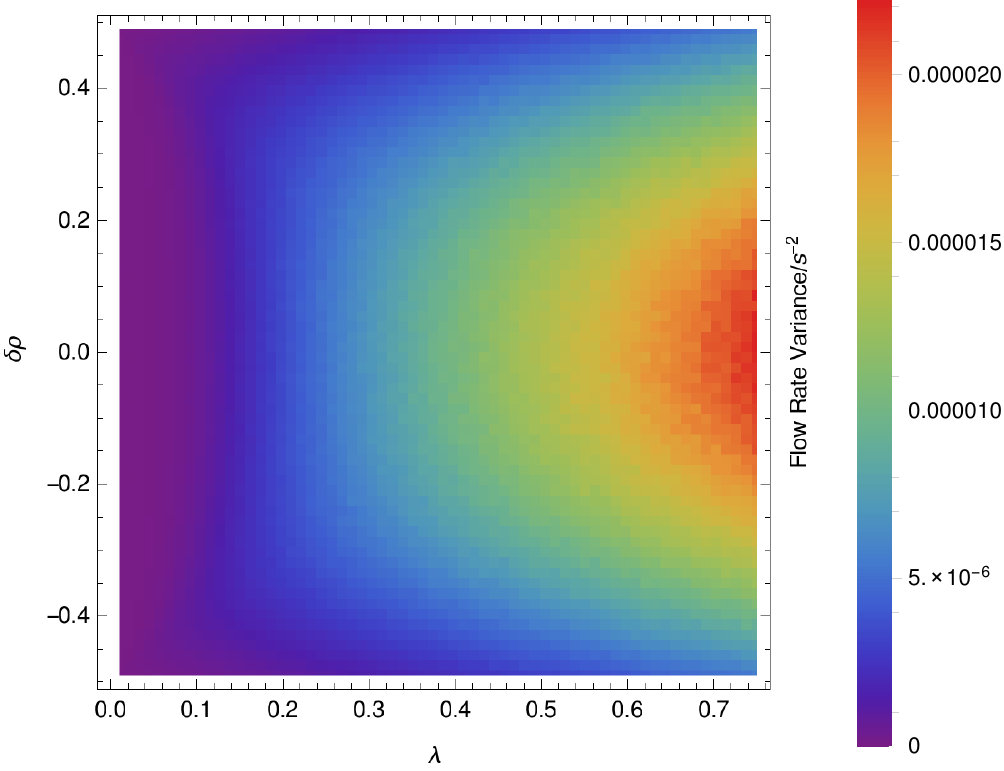} & \includegraphics[width=0.5\linewidth]{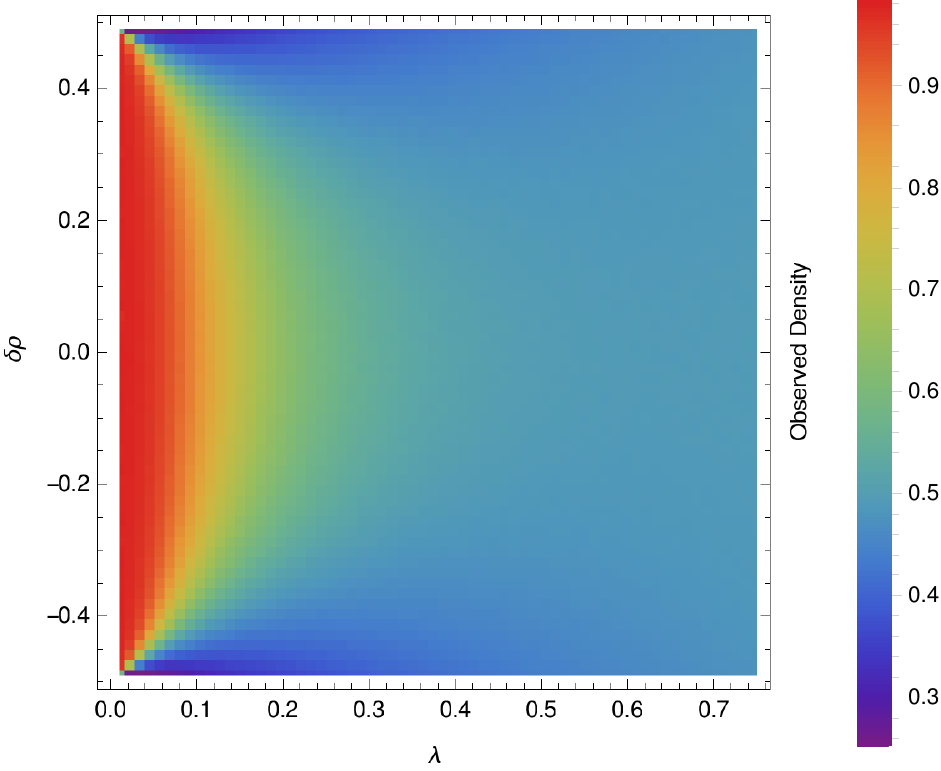} \\
    \end{tabular}
\end{center}
    \vspace{0em}
\end{figure*}

\begin{figure*}[h!]
\vspace{1em}
\caption{\label{fig:fullDiffCoef}
These images are in addition to our computation of the effective diffusion coefficient $D$. The left shows the overall system density as a function of $\rho_M$ and $\lambda$, whilst the right shows the standard error in our estimate of $D$.
In this setup we ran the KMC simulation for
$1.6\times10^8$ equilibration steps, followed by $10$ sets of alternating measurement and relaxation runs, of lengths $8\times10^7$ and $1.6\times10^7$ steps respectively. These results are consistent with calculations performed on smaller
systems, so finite-size effects should be suppressed.
}

\begin{center}
 \begin{tabular}{c@{\hspace{1em}}c}
    \includegraphics[width=0.5\linewidth]{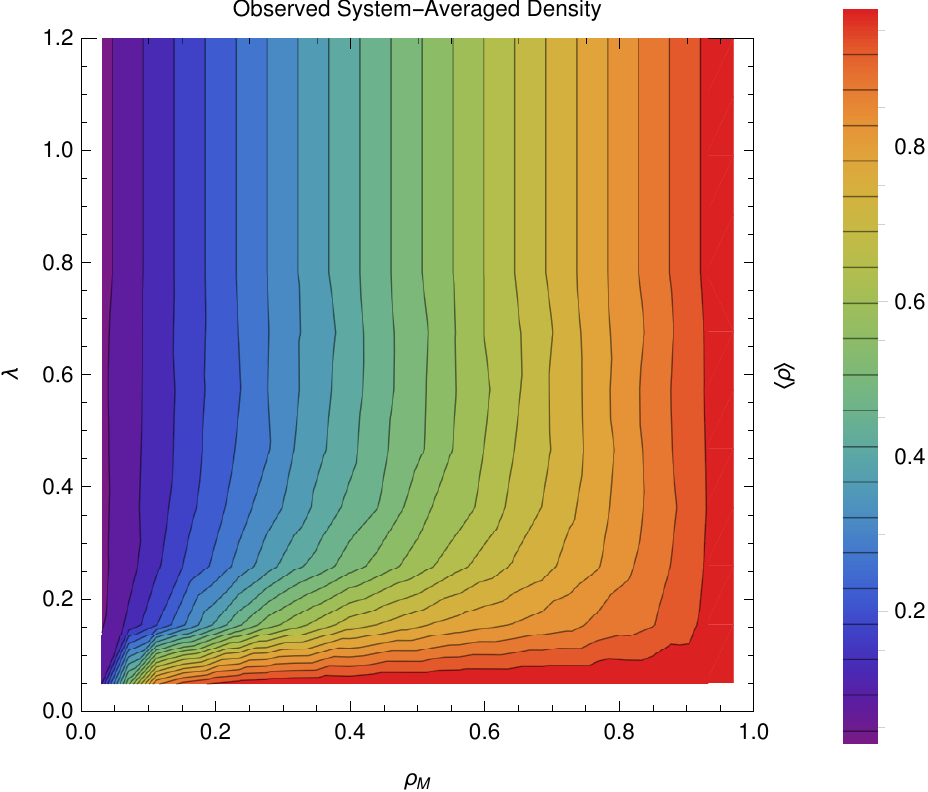} & \includegraphics[width=0.5\linewidth]{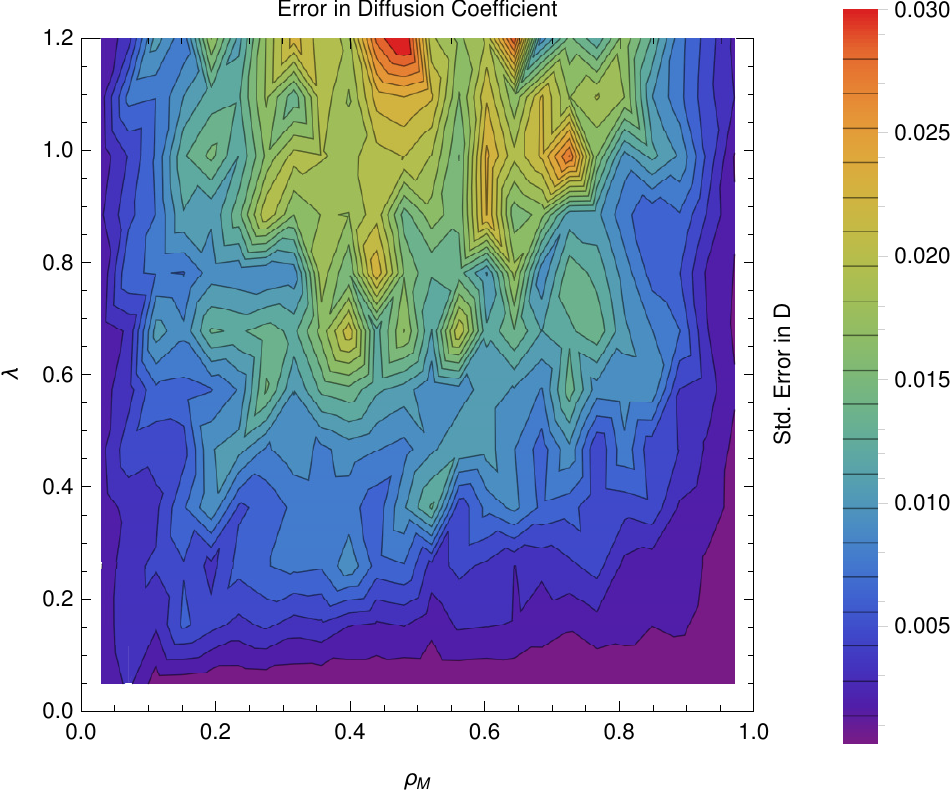} 
    \end{tabular}
\end{center}
    \vspace{0em}
\end{figure*}

Fig.~\ref{fig:fullLambdaScans}, \ref{fig:fullConstDens} and \ref{fig:fullDiffCoef} display additional data and information from computations discussed in the main body of the paper. These were performed with KMC using the same run parameters as those used to create Fig.~\ref{fig:lambdaScans}. 

\section{More flow plots.} For the interested reader we have included for spacetime flow diagrams, show in Fig.~\ref{fig:flowPatterns}. When $\lambda=0.05$, the medium consists of solid blocks surrounded by empty spaces containing a dilute
gas of particles; at $\lambda=0.35$ it is not so different from normal diffusion.
The most interesting images are those for the intermediate $\lambda$; here we see a ``lumpy'' or ``foamy'' structure, in which small blocks
of particles are being constantly created and destroyed whilst a rather minimal flow occurs across the system.
The simulations did not show any hard phase transition as we vary $\lambda$; rather, it seems that this ``foamy''
behavior is part of a continuous range between the extremes, containing medium-range correlations between particles.
Unfortunately, computing equal-time correlation functions to the accuracy required
to draw conclusions about these correlations has proven to be extremely difficult, so we cannot find a quantitative description of the foam beyond the flow rate and density data we have discussed in this paper.
In all images in Fig.~\ref{fig:flowPatterns}, long straight segments of white of black can be seen.  The represent coherent motion at a characteristic velocity given by their gradient. There is nothing in the MFT to suggest what this velocity
should be, and it is much smaller than the simulated system's length divided by the elapsed time,  $\frac{L}{T}$, thus it must be an emergent property arising from correlated motion of self-assembled regions of  high- or low-density material.
However, it has again proved difficult to analyze this numerically.

\begin{figure*}[h!]
\caption{\label{fig:flowPatterns} The spacetime flow patterns, for $\lambda$-values of $0.05$, $0.15$, $0.25$ and $0.35$ going clockwise from top left.
In each plot time runs along the $x$-axis, space along the $y$-axis, with the bottom boundary being held with a concentration of $\frac{1}{4}$ and the top at $\frac{3}{4}$. White represents full occupation, black empty, and gray shades partial
occupation. The degree of occupation was calculated by taking the \texttt{KMCLib} record of a particular site's occupation (i.e. the Gillespie times at
which the site changed occupation), assigning $0$ and $1$ to particles and vacancies respectively, linearly interpolating this and then integrating over times longer than a single Gillespie step but much shorter than the total time in question.
In each case the total time elapsed is that taken by $2^6$ Gillespie steps, and each short-time-average has been done over the total time divided by $512$, the number of lattice sites; this means that on average a site would change state 8 times
per pixel, although of course the real distribution is not nearly this even.
Time has been rescaled this way in order to allow fair comparison of different $\lambda$-values.}
\begin{center}
 \begin{tabular}{c@{\hspace{0.35em}}c}
    \includegraphics[height=0.3\paperheight]{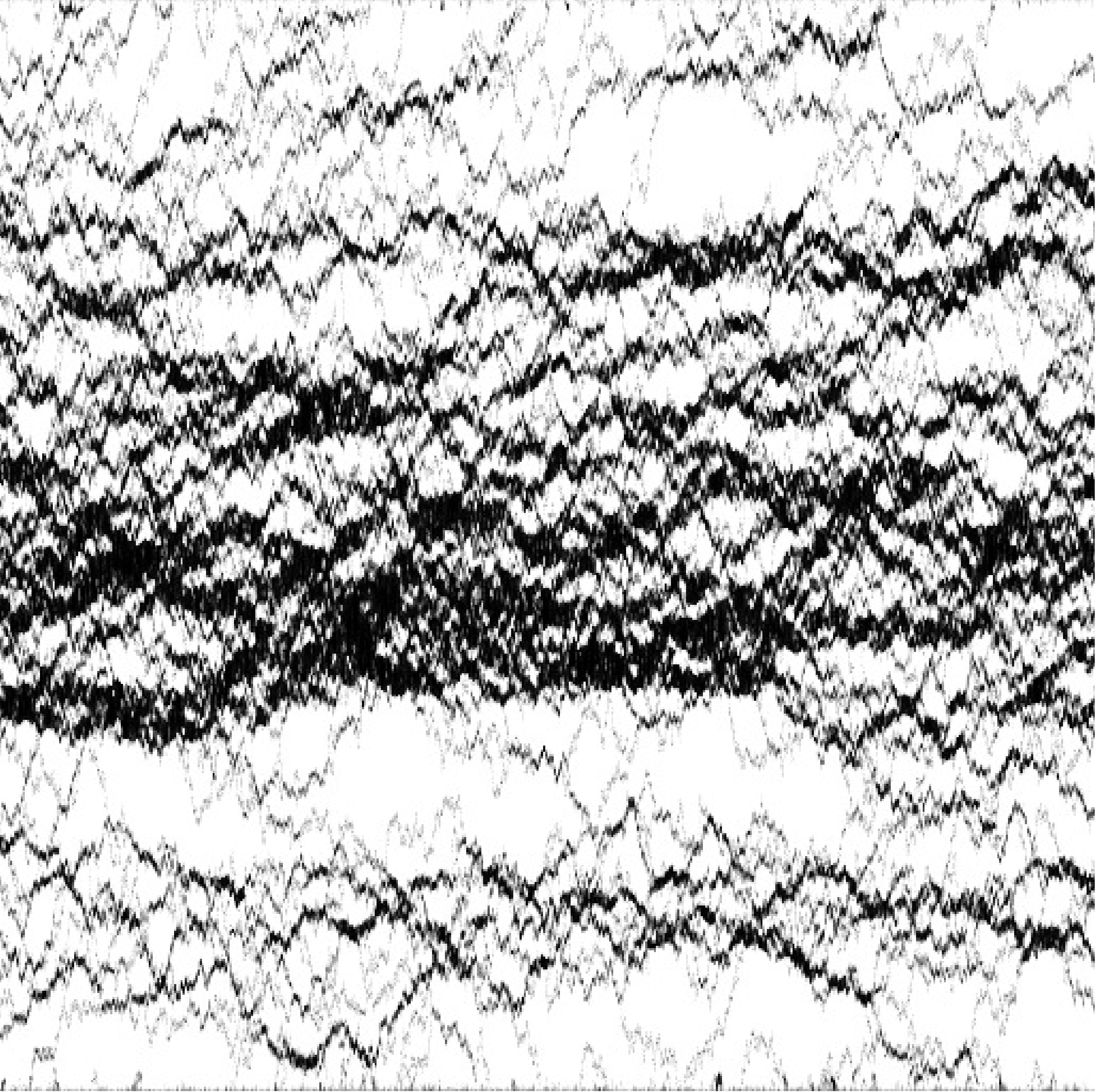} & \includegraphics[height=0.3\paperheight]{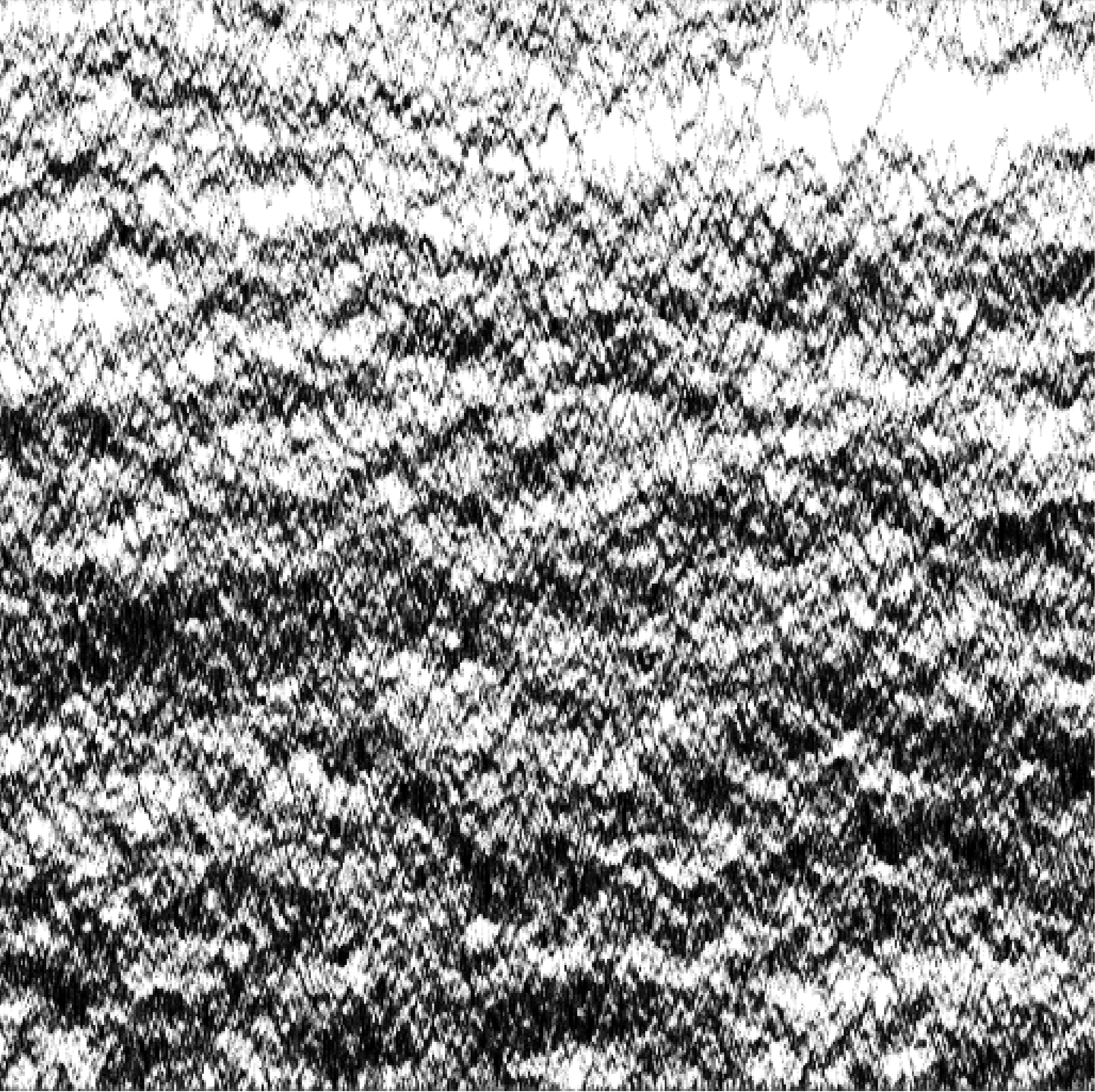} \\
    \includegraphics[height=0.3\paperheight]{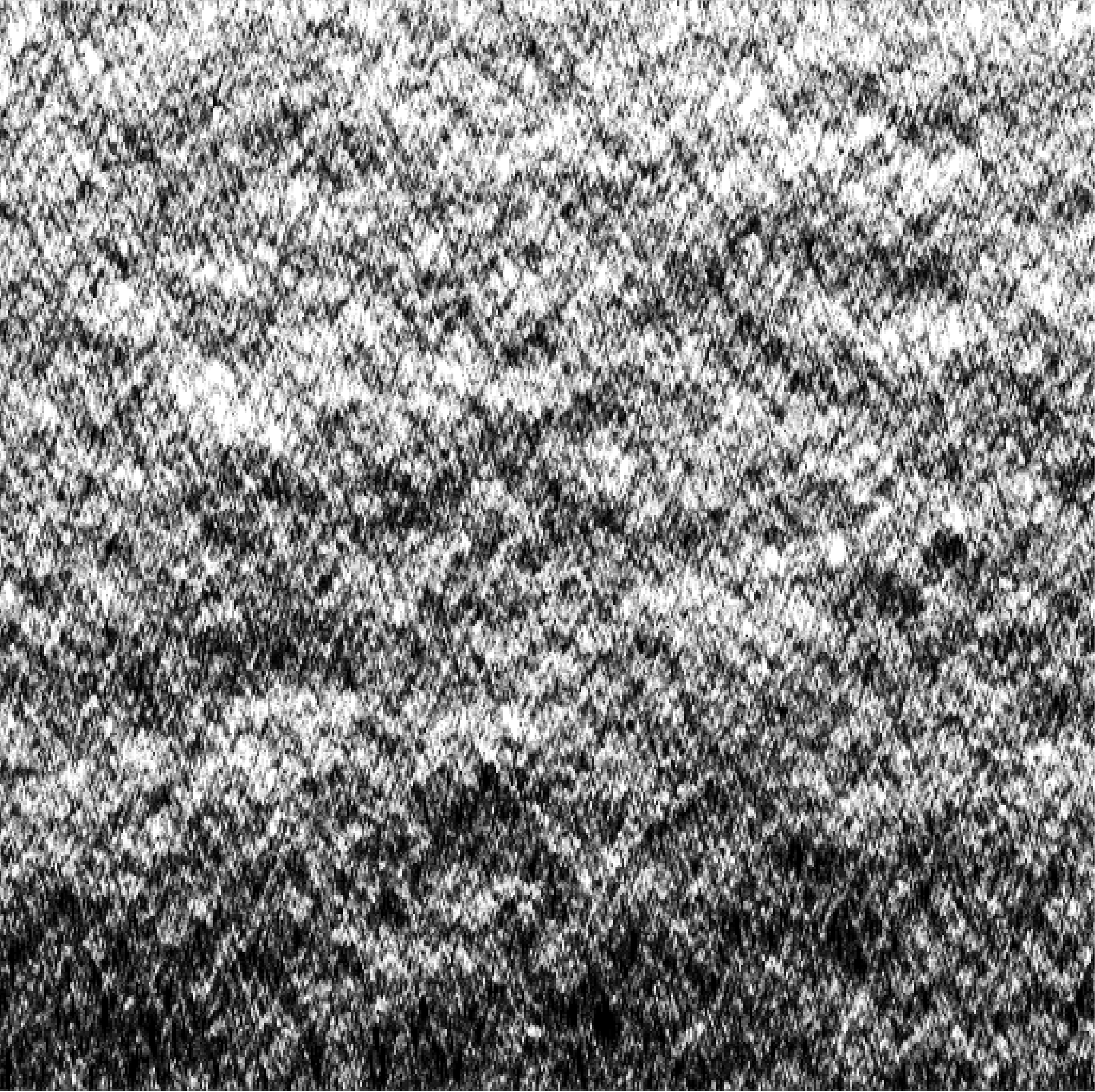} & \includegraphics[height=0.3\paperheight]{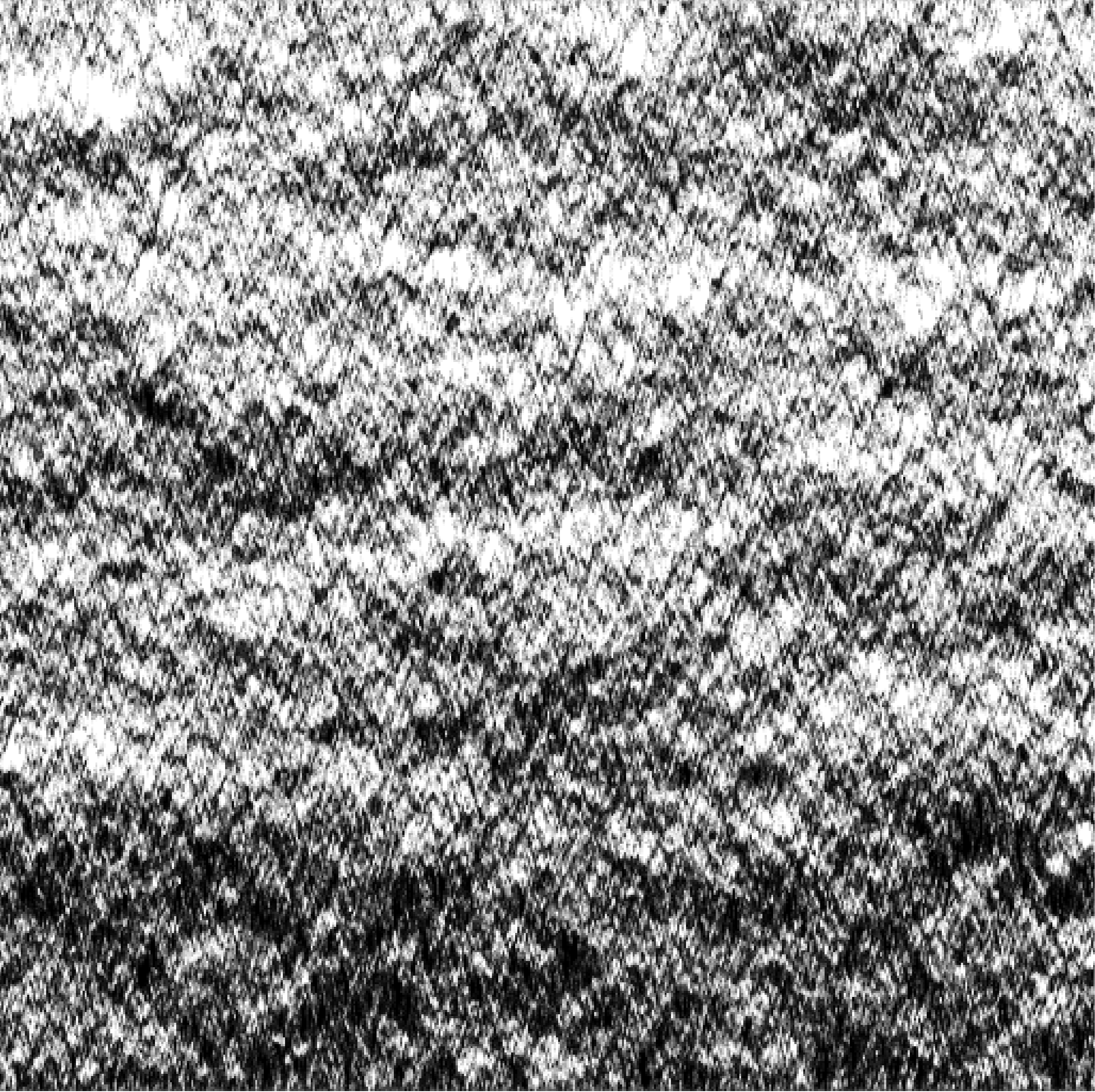} \\
    \end{tabular}
\end{center}
    \vspace{0em}
\end{figure*}

\section{Non-Uniqueness of the MFT Density Profile.}
Recall that, according to the MFT,
\begin{equation}
 \int \! \!  \mathrm{d}x \ J(x) = (x-x_0)J_0 = -\frac{a^2}{\tau_0} \rho \left[1+\zeta \rho\left(\rho-2\right)\right], \label{cubic}
\end{equation}
which we would like to solve for $\rho(x)$. We can do this uniquely so long as the right hand side is monotonic for $\rho \in (0, 1)$. Monotonicity requires that the sign of the derivative of the RHS wrt $\rho$ does not change in that region.
This derivative is of course
\begin{align}
 \frac{\mathrm{d} \mathrm{RHS}}{\mathrm{d} \rho} &= -\frac{a^2}{\tau_0} \rho \left[1+\zeta \rho\left(\rho-2\right)\right] \\
 &= -\frac{a^2}{\tau_0} \left[ 3 \zeta (\rho-\frac{2}{3})^2 + (1-\frac{4}{3} \zeta) \right].
\end{align}
The term in square brackets is always $1$ at $\rho=0$ and $\lambda$ at $\rho=1$, which are both positive.
If $\zeta<0$, the term is an n-shaped parabola with both boundaries positive; therefore it is always positive, and there is no sign change.
For $\zeta>0$ (the case of interest for us), we can see from completing the square that the term is always positive unless $\zeta > \frac{3}{4}$, in which case there is a sign change and therefore a loss of monotonicity.
Thus for $\zeta > \frac{3}{4}$ (and correspondingly $\lambda<\frac{1}{4}$) the MFT does not have a unique steady state solution given a set of Dirichlet boundary conditions, and so we cannot expect the MFT to predict the density profile
in that regime.

\section{Particle Density in Bounded Domain at Extreme $\lambda$-Values.}
\begin{figure*}[h!]
\vspace{1em}
\caption{\label{fig:largeFlow} Flow rate as a function of $\lambda$ for systems with boundary conditions $(\frac{3}{10}, \frac{1}{10})$, $(\frac{3}{4}, \frac{1}{4})$ and $(\frac{9}{10}, \frac{7}{10})$, colored in blue, gold and green
respectively.}
    \includegraphics[width=0.95\linewidth]{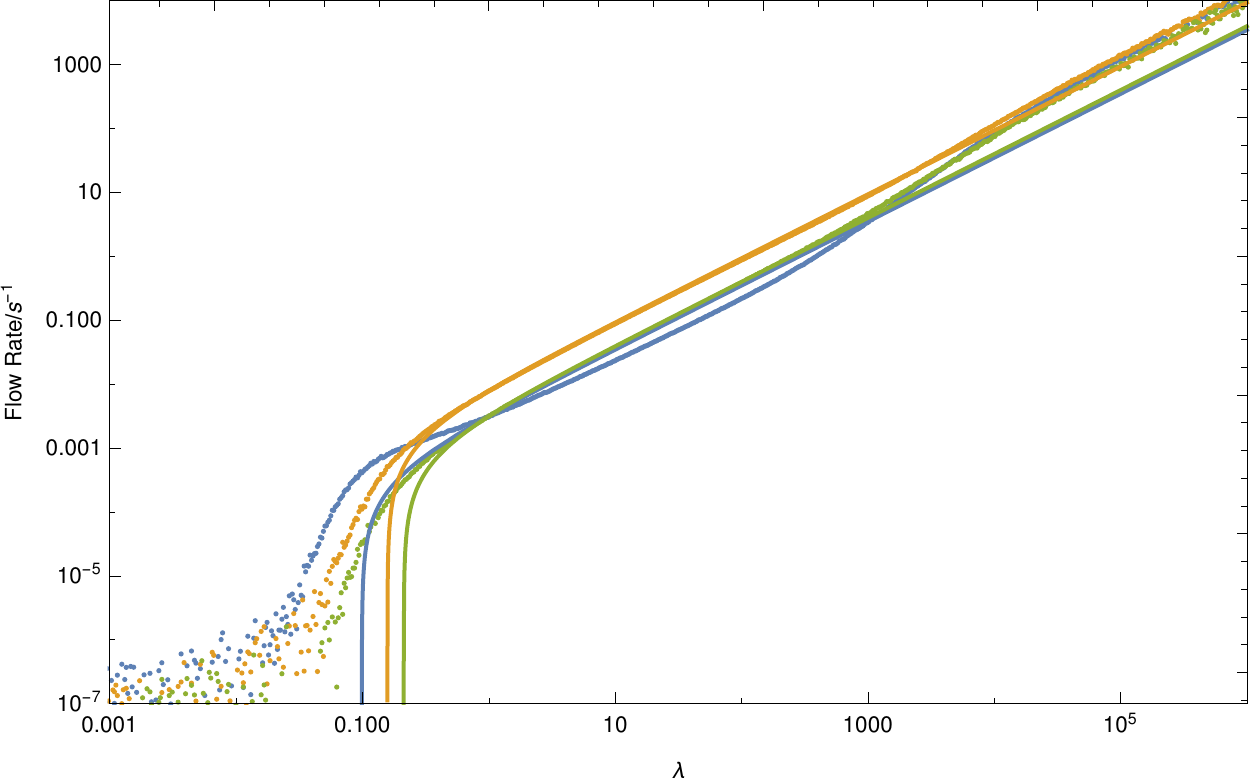}
    \vspace{0em}
\end{figure*}

We scanned across a wide range of $\lambda$ with three sets of boundary conditions: $(\frac{3}{10}, \frac{1}{10})$, $(\frac{3}{4}, \frac{1}{4})$ and $(\frac{9}{10}, \frac{7}{10})$. The resulting mean flow rate is shown in
Fig.~\ref{fig:largeFlow}, and the corresponding density results were already displayed. We once again see the transitions between power law behaviors, as discussed in the main body of the paper. Note that the MFT is never a particularly good
fit for the $(\frac{3}{4}, \frac{1}{4})$ configuration; this may be because the difference between the boundaries is greater than in the other cases. The other MFTs are good fits in the high-$\lambda$ regime until we start reaching $\lambda~\sim~1000$,
at which point they seem to start converging to the same flow regardless of boundary conditions. In each case for low-$\lambda$ we have a power-law regime, each with with an exponent around $4$, and then at extreme low-$\lambda$ we lose
a consistent signal because of noise, which we attribute to lack of system convergence due to critically slow flow.
So far as the density is concerned, at $\lambda=1$ in each case it is about what we would expect (the average of the two boundary densities). It is minimized around $\lambda = 0.6$, and then as $\lambda$ is reduced each density appears to 
converge to $1$ regardless of boundary conditions (before we encounter the same convergence issue at extreme low-$\lambda$). As we go to high-$\lambda$, on the other hand, we see a similar convergence, but this time towards $\rho \sim 0.7$.
This is interesting, as a density of $\rho=\frac{2}{3}$ is predicted by the MFT to be where maximal flow should occur for $\lambda>1$; thus it would appear that the system self-organizes to facilitate maximal flow, which is something which
is hypothesized to occur in other nonequilibrium statmech systems. This also helps to explain the large deviations from the MFT flow predictions we encounter at high-$\lambda$, as the system in practice has a different density to the one assumed
by the MFT.


\begin{thebibliography}{37}%
\makeatletter
\providecommand \@ifxundefined [1]{%
 \@ifx{#1\undefined}
}%
\providecommand \@ifnum [1]{%
 \ifnum #1\expandafter \@firstoftwo
 \else \expandafter \@secondoftwo
 \fi
}%
\providecommand \@ifx [1]{%
 \ifx #1\expandafter \@firstoftwo
 \else \expandafter \@secondoftwo
 \fi
}%
\providecommand \natexlab [1]{#1}%
\providecommand \enquote  [1]{``#1''}%
\providecommand \bibnamefont  [1]{#1}%
\providecommand \bibfnamefont [1]{#1}%
\providecommand \citenamefont [1]{#1}%
\providecommand \href@noop [0]{\@secondoftwo}%
\providecommand \href [0]{\begingroup \@sanitize@url \@href}%
\providecommand \@href[1]{\@@startlink{#1}\@@href}%
\providecommand \@@href[1]{\endgroup#1\@@endlink}%
\providecommand \@sanitize@url [0]{\catcode `\\12\catcode `\$12\catcode
  `\&12\catcode `\#12\catcode `\^12\catcode `\_12\catcode `\%12\relax}%
\providecommand \@@startlink[1]{}%
\providecommand \@@endlink[0]{}%
\providecommand \url  [0]{\begingroup\@sanitize@url \@url }%
\providecommand \@url [1]{\endgroup\@href {#1}{\urlprefix }}%
\providecommand \urlprefix  [0]{URL }%
\providecommand \Eprint [0]{\href }%
\providecommand \doibase [0]{http://dx.doi.org/}%
\providecommand \selectlanguage [0]{\@gobble}%
\providecommand \bibinfo  [0]{\@secondoftwo}%
\providecommand \bibfield  [0]{\@secondoftwo}%
\providecommand \translation [1]{[#1]}%
\providecommand \BibitemOpen [0]{}%
\providecommand \bibitemStop [0]{}%
\providecommand \bibitemNoStop [0]{.\EOS\space}%
\providecommand \EOS [0]{\spacefactor3000\relax}%
\providecommand \BibitemShut  [1]{\csname bibitem#1\endcsname}%
\let\auto@bib@innerbib\@empty
\bibitem [{\citenamefont {Belitsky}\ and\ \citenamefont
  {Sch{\"u}tz}(2011)}]{1742-5468-2011-07-P07007}%
  \BibitemOpen
  \bibfield  {author} {\bibinfo {author} {\bibfnamefont {V}~\bibnamefont
  {Belitsky}}\ and\ \bibinfo {author} {\bibfnamefont {G~M}\ \bibnamefont
  {Sch{\"u}tz}},\ }\bibfield  {title} {\enquote {\bibinfo {title} {Cellular
  automaton model for molecular traffic jams},}\ }\href
  {http://stacks.iop.org/1742-5468/2011/i=07/a=P07007} {\bibfield  {journal}
  {\bibinfo  {journal} {Journal of Statistical Mechanics: Theory and
  Experiment}\ }\textbf {\bibinfo {volume} {2011}},\ \bibinfo {pages} {P07007}
  (\bibinfo {year} {2011})}\BibitemShut {NoStop}%
\bibitem [{\citenamefont {Mobilia}\ \emph {et~al.}(2007)\citenamefont
  {Mobilia}, \citenamefont {Georgiev},\ and\ \citenamefont
  {T{\"a}uber}}]{Mobilia2007}%
  \BibitemOpen
  \bibfield  {author} {\bibinfo {author} {\bibfnamefont {Mauro}\ \bibnamefont
  {Mobilia}}, \bibinfo {author} {\bibfnamefont {Ivan~T.}\ \bibnamefont
  {Georgiev}}, \ and\ \bibinfo {author} {\bibfnamefont {Uwe~C.}\ \bibnamefont
  {T{\"a}uber}},\ }\bibfield  {title} {\enquote {\bibinfo {title} {Phase
  transitions and spatio-temporal fluctuations in stochastic lattice
  lotka--volterra models},}\ }\href {\doibase 10.1007/s10955-006-9146-3}
  {\bibfield  {journal} {\bibinfo  {journal} {Journal of Statistical Physics}\
  }\textbf {\bibinfo {volume} {128}},\ \bibinfo {pages} {447--483} (\bibinfo
  {year} {2007})}\BibitemShut {NoStop}%
\bibitem [{\citenamefont {Tegner}\ \emph {et~al.}(2015)\citenamefont {Tegner},
  \citenamefont {Zhu}, \citenamefont {Siemers}, \citenamefont {Saksl},\ and\
  \citenamefont {Ackland}}]{tegner2015high}%
  \BibitemOpen
  \bibfield  {author} {\bibinfo {author} {\bibfnamefont {{B. E.}}\ \bibnamefont
  {Tegner}}, \bibinfo {author} {\bibfnamefont {L.}~\bibnamefont {Zhu}},
  \bibinfo {author} {\bibfnamefont {C.}~\bibnamefont {Siemers}}, \bibinfo
  {author} {\bibfnamefont {K.}~\bibnamefont {Saksl}}, \ and\ \bibinfo {author}
  {\bibfnamefont {{G. J.}}\ \bibnamefont {Ackland}},\ }\bibfield  {title}
  {\enquote {\bibinfo {title} {High temperature oxidation resistance in
  titanium-niobium alloys},}\ }\href {\doibase 10.1016/j.jallcom.2015.04.115}
  {\bibfield  {journal} {\bibinfo  {journal} {Journal of alloys and compounds}\
  }\textbf {\bibinfo {volume} {643}},\ \bibinfo {pages} {100--105} (\bibinfo
  {year} {2015})}\BibitemShut {NoStop}%
\bibitem [{\citenamefont {Zhu}\ \emph {et~al.}(2012)\citenamefont {Zhu},
  \citenamefont {Hu}, \citenamefont {Yang},\ and\ \citenamefont
  {Ackland}}]{zhu2012atomic}%
  \BibitemOpen
  \bibfield  {author} {\bibinfo {author} {\bibfnamefont {Linggang}\
  \bibnamefont {Zhu}}, \bibinfo {author} {\bibfnamefont {Qing-Miao}\
  \bibnamefont {Hu}}, \bibinfo {author} {\bibfnamefont {Rui}\ \bibnamefont
  {Yang}}, \ and\ \bibinfo {author} {\bibfnamefont {{Graeme J.}}\ \bibnamefont
  {Ackland}},\ }\bibfield  {title} {\enquote {\bibinfo {title} {Atomic-scale
  modeling of the dynamics of titanium oxidation},}\ }\href {\doibase
  10.1021/jp309305n} {\bibfield  {journal} {\bibinfo  {journal} {Journal of
  Physical Chemistry C}\ }\textbf {\bibinfo {volume} {116}},\ \bibinfo {pages}
  {24201--24205} (\bibinfo {year} {2012})}\BibitemShut {NoStop}%
\bibitem [{\citenamefont {Deal}\ and\ \citenamefont
  {Grove}(1965)}]{DealGrove1965}%
  \BibitemOpen
  \bibfield  {author} {\bibinfo {author} {\bibfnamefont {B.~E.}\ \bibnamefont
  {Deal}}\ and\ \bibinfo {author} {\bibfnamefont {A.~S.}\ \bibnamefont
  {Grove}},\ }\bibfield  {title} {\enquote {\bibinfo {title} {General
  relationship for the thermal oxidation of silicon},}\ }\href {\doibase
  10.1063/1.1713945} {\bibfield  {journal} {\bibinfo  {journal} {Journal of
  Applied Physics}\ }\textbf {\bibinfo {volume} {36}},\ \bibinfo {pages}
  {3770--3778} (\bibinfo {year} {1965})},\ \Eprint
  {http://arxiv.org/abs/https://doi.org/10.1063/1.1713945}
  {https://doi.org/10.1063/1.1713945} \BibitemShut {NoStop}%
\bibitem [{\citenamefont {Cabrera}\ and\ \citenamefont
  {Mott}(1949)}]{MottCabrera1949}%
  \BibitemOpen
  \bibfield  {author} {\bibinfo {author} {\bibfnamefont {N}~\bibnamefont
  {Cabrera}}\ and\ \bibinfo {author} {\bibfnamefont {N~F}\ \bibnamefont
  {Mott}},\ }\bibfield  {title} {\enquote {\bibinfo {title} {Theory of the
  oxidation of metals},}\ }\href {http://stacks.iop.org/0034-4885/12/i=1/a=308}
  {\bibfield  {journal} {\bibinfo  {journal} {Reports on Progress in Physics}\
  }\textbf {\bibinfo {volume} {12}},\ \bibinfo {pages} {163} (\bibinfo {year}
  {1949})}\BibitemShut {NoStop}%
\bibitem [{\citenamefont {Buzzaccaro}\ \emph {et~al.}(2007)\citenamefont
  {Buzzaccaro}, \citenamefont {Rusconi},\ and\ \citenamefont
  {Piazza}}]{Buzzaccaro2007}%
  \BibitemOpen
  \bibfield  {author} {\bibinfo {author} {\bibfnamefont {Stefano}\ \bibnamefont
  {Buzzaccaro}}, \bibinfo {author} {\bibfnamefont {Roberto}\ \bibnamefont
  {Rusconi}}, \ and\ \bibinfo {author} {\bibfnamefont {Roberto}\ \bibnamefont
  {Piazza}},\ }\bibfield  {title} {\enquote {\bibinfo {title} {``sticky'' hard
  spheres: Equation of state, phase diagram, and metastable gels},}\ }\href
  {\doibase 10.1103/PhysRevLett.99.098301} {\bibfield  {journal} {\bibinfo
  {journal} {Phys. Rev. Lett.}\ }\textbf {\bibinfo {volume} {99}},\ \bibinfo
  {pages} {098301} (\bibinfo {year} {2007})}\BibitemShut {NoStop}%
\bibitem [{\citenamefont {Ladd}\ \emph {et~al.}(1988)\citenamefont {Ladd},
  \citenamefont {Colvin},\ and\ \citenamefont {Frenkel}}]{ladd1988application}%
  \BibitemOpen
  \bibfield  {author} {\bibinfo {author} {\bibfnamefont {Anthony J.~C.}\
  \bibnamefont {Ladd}}, \bibinfo {author} {\bibfnamefont {Michael~E.}\
  \bibnamefont {Colvin}}, \ and\ \bibinfo {author} {\bibfnamefont {Daan}\
  \bibnamefont {Frenkel}},\ }\bibfield  {title} {\enquote {\bibinfo {title}
  {Application of lattice-gas cellular automata to the brownian motion of
  solids in suspension},}\ }\href {\doibase 10.1103/PhysRevLett.60.975}
  {\bibfield  {journal} {\bibinfo  {journal} {Phys. Rev. Lett.}\ }\textbf
  {\bibinfo {volume} {60}},\ \bibinfo {pages} {975--978} (\bibinfo {year}
  {1988})}\BibitemShut {NoStop}%
\bibitem [{\citenamefont {Liggett}(1985)}]{liggett1985interacting}%
  \BibitemOpen
  \bibfield  {author} {\bibinfo {author} {\bibfnamefont {Thomas~M}\
  \bibnamefont {Liggett}},\ }\href@noop {} {\emph {\bibinfo {title}
  {Interacting particle systems}}}\ (\bibinfo  {publisher} {Springer-Verlag,
  Berlin},\ \bibinfo {year} {1985})\BibitemShut {NoStop}%
\bibitem [{\citenamefont {Ben-Naim}\ \emph {et~al.}(1999)\citenamefont
  {Ben-Naim}, \citenamefont {Chen}, \citenamefont {Doolen},\ and\ \citenamefont
  {Redner}}]{BenNaim1999}%
  \BibitemOpen
  \bibfield  {author} {\bibinfo {author} {\bibfnamefont {E.}~\bibnamefont
  {Ben-Naim}}, \bibinfo {author} {\bibfnamefont {S.~Y.}\ \bibnamefont {Chen}},
  \bibinfo {author} {\bibfnamefont {G.~D.}\ \bibnamefont {Doolen}}, \ and\
  \bibinfo {author} {\bibfnamefont {S.}~\bibnamefont {Redner}},\ }\bibfield
  {title} {\enquote {\bibinfo {title} {Shocklike dynamics of inelastic
  gases},}\ }\href {\doibase 10.1103/PhysRevLett.83.4069} {\bibfield  {journal}
  {\bibinfo  {journal} {Phys. Rev. Lett.}\ }\textbf {\bibinfo {volume} {83}},\
  \bibinfo {pages} {4069--4072} (\bibinfo {year} {1999})}\BibitemShut {NoStop}%
\bibitem [{\citenamefont {Shandarin}\ and\ \citenamefont
  {Zeldovich}(1989)}]{Shandarin1989}%
  \BibitemOpen
  \bibfield  {author} {\bibinfo {author} {\bibfnamefont {S.~F.}\ \bibnamefont
  {Shandarin}}\ and\ \bibinfo {author} {\bibfnamefont {Ya.~B.}\ \bibnamefont
  {Zeldovich}},\ }\bibfield  {title} {\enquote {\bibinfo {title} {The
  large-scale structure of the universe: Turbulence, intermittency, structures
  in a self-gravitating medium},}\ }\href {\doibase 10.1103/RevModPhys.61.185}
  {\bibfield  {journal} {\bibinfo  {journal} {Rev. Mod. Phys.}\ }\textbf
  {\bibinfo {volume} {61}},\ \bibinfo {pages} {185--220} (\bibinfo {year}
  {1989})}\BibitemShut {NoStop}%
\bibitem [{\citenamefont {Frachebourg}(1999)}]{Frachebourg1999}%
  \BibitemOpen
  \bibfield  {author} {\bibinfo {author} {\bibfnamefont {L.}~\bibnamefont
  {Frachebourg}},\ }\bibfield  {title} {\enquote {\bibinfo {title} {Exact
  solution of the one-dimensional ballistic aggregation},}\ }\href {\doibase
  10.1103/PhysRevLett.82.1502} {\bibfield  {journal} {\bibinfo  {journal}
  {Phys. Rev. Lett.}\ }\textbf {\bibinfo {volume} {82}},\ \bibinfo {pages}
  {1502--1505} (\bibinfo {year} {1999})}\BibitemShut {NoStop}%
\bibitem [{\citenamefont {{Frachebourg}}\ \emph {et~al.}(2000)\citenamefont
  {{Frachebourg}}, \citenamefont {{Martin}},\ and\ \citenamefont
  {{Piasecki}}}]{Frachebourg2000}%
  \BibitemOpen
  \bibfield  {author} {\bibinfo {author} {\bibfnamefont {L.}~\bibnamefont
  {{Frachebourg}}}, \bibinfo {author} {\bibfnamefont {P.~A.}\ \bibnamefont
  {{Martin}}}, \ and\ \bibinfo {author} {\bibfnamefont {J.}~\bibnamefont
  {{Piasecki}}},\ }\bibfield  {title} {\enquote {\bibinfo {title} {{Ballistic
  aggregation: a solvable model of irreversible many particles dynamics}},}\
  }\href {\doibase 10.1016/S0378-4371(99)00585-3} {\bibfield  {journal}
  {\bibinfo  {journal} {Physica A Statistical Mechanics and its Applications}\
  }\textbf {\bibinfo {volume} {279}},\ \bibinfo {pages} {69--99} (\bibinfo
  {year} {2000})},\ \Eprint {http://arxiv.org/abs/cond-mat/9911346}
  {cond-mat/9911346} \BibitemShut {NoStop}%
\bibitem [{\citenamefont {Obukhovsky}\ \emph {et~al.}(2017)\citenamefont
  {Obukhovsky}, \citenamefont {Kutsyk}, \citenamefont {Nikonova},\ and\
  \citenamefont {Ilchenko}}]{Obukhovsky2017}%
  \BibitemOpen
  \bibfield  {author} {\bibinfo {author} {\bibfnamefont {Vyacheslav~V.}\
  \bibnamefont {Obukhovsky}}, \bibinfo {author} {\bibfnamefont {Andrii~M.}\
  \bibnamefont {Kutsyk}}, \bibinfo {author} {\bibfnamefont {Viktoria~V.}\
  \bibnamefont {Nikonova}}, \ and\ \bibinfo {author} {\bibfnamefont
  {Oleksii~O.}\ \bibnamefont {Ilchenko}},\ }\bibfield  {title} {\enquote
  {\bibinfo {title} {Nonlinear diffusion in multicomponent liquid solutions},}\
  }\href {\doibase 10.1103/PhysRevE.95.022133} {\bibfield  {journal} {\bibinfo
  {journal} {Phys. Rev. E}\ }\textbf {\bibinfo {volume} {95}},\ \bibinfo
  {pages} {022133} (\bibinfo {year} {2017})}\BibitemShut {NoStop}%
\bibitem [{\citenamefont {Gorokhova}\ and\ \citenamefont
  {Melnik}(2010)}]{Gorokhova2010}%
  \BibitemOpen
  \bibfield  {author} {\bibinfo {author} {\bibfnamefont {N.~V.}\ \bibnamefont
  {Gorokhova}}\ and\ \bibinfo {author} {\bibfnamefont {O.~E.}\ \bibnamefont
  {Melnik}},\ }\bibfield  {title} {\enquote {\bibinfo {title} {Modeling of the
  dynamics of diffusion crystal growth from a cooling magmatic melt},}\ }\href
  {\doibase 10.1134/S0015462810050017} {\bibfield  {journal} {\bibinfo
  {journal} {Fluid Dynamics}\ }\textbf {\bibinfo {volume} {45}},\ \bibinfo
  {pages} {679--690} (\bibinfo {year} {2010})}\BibitemShut {NoStop}%
\bibitem [{\citenamefont {Kardar}\ \emph {et~al.}(1986)\citenamefont {Kardar},
  \citenamefont {Parisi},\ and\ \citenamefont {Zhang}}]{PhysRevLett.56.889}%
  \BibitemOpen
  \bibfield  {author} {\bibinfo {author} {\bibfnamefont {Mehran}\ \bibnamefont
  {Kardar}}, \bibinfo {author} {\bibfnamefont {Giorgio}\ \bibnamefont
  {Parisi}}, \ and\ \bibinfo {author} {\bibfnamefont {Yi-Cheng}\ \bibnamefont
  {Zhang}},\ }\bibfield  {title} {\enquote {\bibinfo {title} {Dynamic scaling
  of growing interfaces},}\ }\href {\doibase 10.1103/PhysRevLett.56.889}
  {\bibfield  {journal} {\bibinfo  {journal} {Phys. Rev. Lett.}\ }\textbf
  {\bibinfo {volume} {56}},\ \bibinfo {pages} {889--892} (\bibinfo {year}
  {1986})}\BibitemShut {NoStop}%
\bibitem [{\citenamefont {Krug}\ and\ \citenamefont
  {Spohn}(1988)}]{PhysRevA.38.4271}%
  \BibitemOpen
  \bibfield  {author} {\bibinfo {author} {\bibfnamefont {J.}~\bibnamefont
  {Krug}}\ and\ \bibinfo {author} {\bibfnamefont {H.}~\bibnamefont {Spohn}},\
  }\bibfield  {title} {\enquote {\bibinfo {title} {Universality classes for
  deterministic surface growth},}\ }\href {\doibase 10.1103/PhysRevA.38.4271}
  {\bibfield  {journal} {\bibinfo  {journal} {Phys. Rev. A}\ }\textbf {\bibinfo
  {volume} {38}},\ \bibinfo {pages} {4271--4283} (\bibinfo {year}
  {1988})}\BibitemShut {NoStop}%
\bibitem [{\citenamefont {Sasamoto}\ and\ \citenamefont
  {Spohn}(2010)}]{Sasamoto2010}%
  \BibitemOpen
  \bibfield  {author} {\bibinfo {author} {\bibfnamefont {Tomohiro}\
  \bibnamefont {Sasamoto}}\ and\ \bibinfo {author} {\bibfnamefont {Herbert}\
  \bibnamefont {Spohn}},\ }\bibfield  {title} {\enquote {\bibinfo {title}
  {One-dimensional kardar-parisi-zhang equation: An exact solution and its
  universality},}\ }\href {\doibase 10.1103/PhysRevLett.104.230602} {\bibfield
  {journal} {\bibinfo  {journal} {Phys. Rev. Lett.}\ }\textbf {\bibinfo
  {volume} {104}},\ \bibinfo {pages} {230602} (\bibinfo {year}
  {2010})}\BibitemShut {NoStop}%
\bibitem [{\citenamefont {Sugden}\ and\ \citenamefont
  {Evans}(2007)}]{sugden2007dynamically}%
  \BibitemOpen
  \bibfield  {author} {\bibinfo {author} {\bibfnamefont {K~E~P}\ \bibnamefont
  {Sugden}}\ and\ \bibinfo {author} {\bibfnamefont {M~R}\ \bibnamefont
  {Evans}},\ }\bibfield  {title} {\enquote {\bibinfo {title} {A dynamically
  extending exclusion process},}\ }\href
  {http://stacks.iop.org/1742-5468/2007/i=11/a=P11013} {\bibfield  {journal}
  {\bibinfo  {journal} {Journal of Statistical Mechanics: Theory and
  Experiment}\ }\textbf {\bibinfo {volume} {2007}},\ \bibinfo {pages} {P11013}
  (\bibinfo {year} {2007})}\BibitemShut {NoStop}%
\bibitem [{\citenamefont {Kollmann}(2003)}]{Kollmann2003}%
  \BibitemOpen
  \bibfield  {author} {\bibinfo {author} {\bibfnamefont {Markus}\ \bibnamefont
  {Kollmann}},\ }\bibfield  {title} {\enquote {\bibinfo {title} {Single-file
  diffusion of atomic and colloidal systems: Asymptotic laws},}\ }\href
  {\doibase 10.1103/PhysRevLett.90.180602} {\bibfield  {journal} {\bibinfo
  {journal} {Phys. Rev. Lett.}\ }\textbf {\bibinfo {volume} {90}},\ \bibinfo
  {pages} {180602} (\bibinfo {year} {2003})}\BibitemShut {NoStop}%
\bibitem [{\citenamefont {Lin}\ \emph {et~al.}(2005)\citenamefont {Lin},
  \citenamefont {Meron}, \citenamefont {Cui}, \citenamefont {Rice},\ and\
  \citenamefont {Diamant}}]{Lin2005}%
  \BibitemOpen
  \bibfield  {author} {\bibinfo {author} {\bibfnamefont {Binhua}\ \bibnamefont
  {Lin}}, \bibinfo {author} {\bibfnamefont {Mati}\ \bibnamefont {Meron}},
  \bibinfo {author} {\bibfnamefont {Bianxiao}\ \bibnamefont {Cui}}, \bibinfo
  {author} {\bibfnamefont {Stuart~A.}\ \bibnamefont {Rice}}, \ and\ \bibinfo
  {author} {\bibfnamefont {Haim}\ \bibnamefont {Diamant}},\ }\bibfield  {title}
  {\enquote {\bibinfo {title} {From random walk to single-file diffusion},}\
  }\href {\doibase 10.1103/PhysRevLett.94.216001} {\bibfield  {journal}
  {\bibinfo  {journal} {Phys. Rev. Lett.}\ }\textbf {\bibinfo {volume} {94}},\
  \bibinfo {pages} {216001} (\bibinfo {year} {2005})}\BibitemShut {NoStop}%
\bibitem [{\citenamefont {Hegde}\ \emph {et~al.}(2014)\citenamefont {Hegde},
  \citenamefont {Sabhapandit},\ and\ \citenamefont {Dhar}}]{Hegde2014}%
  \BibitemOpen
  \bibfield  {author} {\bibinfo {author} {\bibfnamefont {Chaitra}\ \bibnamefont
  {Hegde}}, \bibinfo {author} {\bibfnamefont {Sanjib}\ \bibnamefont
  {Sabhapandit}}, \ and\ \bibinfo {author} {\bibfnamefont {Abhishek}\
  \bibnamefont {Dhar}},\ }\bibfield  {title} {\enquote {\bibinfo {title}
  {Universal large deviations for the tagged particle in single-file motion},}\
  }\href {\doibase 10.1103/PhysRevLett.113.120601} {\bibfield  {journal}
  {\bibinfo  {journal} {Phys. Rev. Lett.}\ }\textbf {\bibinfo {volume} {113}},\
  \bibinfo {pages} {120601} (\bibinfo {year} {2014})}\BibitemShut {NoStop}%
\bibitem [{\citenamefont {Krapivsky}\ \emph {et~al.}(2014)\citenamefont
  {Krapivsky}, \citenamefont {Mallick},\ and\ \citenamefont
  {Sadhu}}]{Krapivsky2014}%
  \BibitemOpen
  \bibfield  {author} {\bibinfo {author} {\bibfnamefont {P.~L.}\ \bibnamefont
  {Krapivsky}}, \bibinfo {author} {\bibfnamefont {Kirone}\ \bibnamefont
  {Mallick}}, \ and\ \bibinfo {author} {\bibfnamefont {Tridib}\ \bibnamefont
  {Sadhu}},\ }\bibfield  {title} {\enquote {\bibinfo {title} {Large deviations
  in single-file diffusion},}\ }\href {\doibase 10.1103/PhysRevLett.113.078101}
  {\bibfield  {journal} {\bibinfo  {journal} {Phys. Rev. Lett.}\ }\textbf
  {\bibinfo {volume} {113}},\ \bibinfo {pages} {078101} (\bibinfo {year}
  {2014})}\BibitemShut {NoStop}%
\bibitem [{\citenamefont {Imamura}\ \emph {et~al.}(2017)\citenamefont
  {Imamura}, \citenamefont {Mallick},\ and\ \citenamefont
  {Sasamoto}}]{Imamura2017}%
  \BibitemOpen
  \bibfield  {author} {\bibinfo {author} {\bibfnamefont {Takashi}\ \bibnamefont
  {Imamura}}, \bibinfo {author} {\bibfnamefont {Kirone}\ \bibnamefont
  {Mallick}}, \ and\ \bibinfo {author} {\bibfnamefont {Tomohiro}\ \bibnamefont
  {Sasamoto}},\ }\bibfield  {title} {\enquote {\bibinfo {title} {Large
  deviations of a tracer in the symmetric exclusion process},}\ }\href
  {\doibase 10.1103/PhysRevLett.118.160601} {\bibfield  {journal} {\bibinfo
  {journal} {Phys. Rev. Lett.}\ }\textbf {\bibinfo {volume} {118}},\ \bibinfo
  {pages} {160601} (\bibinfo {year} {2017})}\BibitemShut {NoStop}%
\bibitem [{\citenamefont {Katz}\ \emph {et~al.}(1984)\citenamefont {Katz},
  \citenamefont {Lebowitz},\ and\ \citenamefont {Spohn}}]{Katz1984}%
  \BibitemOpen
  \bibfield  {author} {\bibinfo {author} {\bibfnamefont {Sheldon}\ \bibnamefont
  {Katz}}, \bibinfo {author} {\bibfnamefont {Joel~L.}\ \bibnamefont
  {Lebowitz}}, \ and\ \bibinfo {author} {\bibfnamefont {Herbert}\ \bibnamefont
  {Spohn}},\ }\bibfield  {title} {\enquote {\bibinfo {title} {Nonequilibrium
  steady states of stochastic lattice gas models of fast ionic conductors},}\
  }\href {\doibase 10.1007/BF01018556} {\bibfield  {journal} {\bibinfo
  {journal} {Journal of Statistical Physics}\ }\textbf {\bibinfo {volume}
  {34}},\ \bibinfo {pages} {497--537} (\bibinfo {year} {1984})}\BibitemShut
  {NoStop}%
\bibitem [{\citenamefont {Zia}(2010)}]{Zia2010}%
  \BibitemOpen
  \bibfield  {author} {\bibinfo {author} {\bibfnamefont {R.~K.~P.}\
  \bibnamefont {Zia}},\ }\bibfield  {title} {\enquote {\bibinfo {title} {Twenty
  five years after kls: A celebration of non-equilibrium statistical
  mechanics},}\ }\href {\doibase 10.1007/s10955-009-9884-0} {\bibfield
  {journal} {\bibinfo  {journal} {Journal of Statistical Physics}\ }\textbf
  {\bibinfo {volume} {138}},\ \bibinfo {pages} {20--28} (\bibinfo {year}
  {2010})}\BibitemShut {NoStop}%
\bibitem [{\citenamefont {Kafri}\ \emph {et~al.}(2003)\citenamefont {Kafri},
  \citenamefont {Levine}, \citenamefont {Mukamel}, \citenamefont {Sch\"utz},\
  and\ \citenamefont {Willmann}}]{Kafri2003}%
  \BibitemOpen
  \bibfield  {author} {\bibinfo {author} {\bibfnamefont {Y.}~\bibnamefont
  {Kafri}}, \bibinfo {author} {\bibfnamefont {E.}~\bibnamefont {Levine}},
  \bibinfo {author} {\bibfnamefont {D.}~\bibnamefont {Mukamel}}, \bibinfo
  {author} {\bibfnamefont {G.~M.}\ \bibnamefont {Sch\"utz}}, \ and\ \bibinfo
  {author} {\bibfnamefont {R.~D.}\ \bibnamefont {Willmann}},\ }\bibfield
  {title} {\enquote {\bibinfo {title} {Phase-separation transition in
  one-dimensional driven models},}\ }\href {\doibase
  10.1103/PhysRevE.68.035101} {\bibfield  {journal} {\bibinfo  {journal} {Phys.
  Rev. E}\ }\textbf {\bibinfo {volume} {68}},\ \bibinfo {pages} {035101}
  (\bibinfo {year} {2003})}\BibitemShut {NoStop}%
\bibitem [{\citenamefont {Kawasaki}(1966)}]{PhysRev.145.224}%
  \BibitemOpen
  \bibfield  {author} {\bibinfo {author} {\bibfnamefont {Kyozi}\ \bibnamefont
  {Kawasaki}},\ }\bibfield  {title} {\enquote {\bibinfo {title} {Diffusion
  constants near the critical point for time-dependent ising models. i},}\
  }\href {\doibase 10.1103/PhysRev.145.224} {\bibfield  {journal} {\bibinfo
  {journal} {Phys. Rev.}\ }\textbf {\bibinfo {volume} {145}},\ \bibinfo {pages}
  {224--230} (\bibinfo {year} {1966})}\BibitemShut {NoStop}%
\bibitem [{\citenamefont {Spohn}(1983)}]{spohn1983}%
  \BibitemOpen
  \bibfield  {author} {\bibinfo {author} {\bibfnamefont {H}~\bibnamefont
  {Spohn}},\ }\bibfield  {title} {\enquote {\bibinfo {title} {Long range
  correlations for stochastic lattice gases in a non-equilibrium steady
  state},}\ }\href {http://stacks.iop.org/0305-4470/16/i=18/a=029} {\bibfield
  {journal} {\bibinfo  {journal} {Journal of Physics A: Mathematical and
  General}\ }\textbf {\bibinfo {volume} {16}},\ \bibinfo {pages} {4275}
  (\bibinfo {year} {1983})}\BibitemShut {NoStop}%
\bibitem [{\citenamefont {{Leetmaa}}\ and\ \citenamefont
  {{Skorodumova}}(2014)}]{leetmaa2014kmclib}%
  \BibitemOpen
  \bibfield  {author} {\bibinfo {author} {\bibfnamefont {M.}~\bibnamefont
  {{Leetmaa}}}\ and\ \bibinfo {author} {\bibfnamefont {N.~V.}\ \bibnamefont
  {{Skorodumova}}},\ }\bibfield  {title} {\enquote {\bibinfo {title} {{KMCLib:
  A general framework for lattice kinetic Monte Carlo (KMC) simulations}},}\
  }\href {\doibase 10.1016/j.cpc.2014.04.017} {\bibfield  {journal} {\bibinfo
  {journal} {Computer Physics Communications}\ }\textbf {\bibinfo {volume}
  {185}},\ \bibinfo {pages} {2340--2349} (\bibinfo {year} {2014})},\ \Eprint
  {http://arxiv.org/abs/1405.1221} {arXiv:1405.1221 [physics.comp-ph]}
  \BibitemShut {NoStop}%
\bibitem [{\citenamefont {Gillespie}(1977)}]{Gillespie1977}%
  \BibitemOpen
  \bibfield  {author} {\bibinfo {author} {\bibfnamefont {Daniel~T.}\
  \bibnamefont {Gillespie}},\ }\bibfield  {title} {\enquote {\bibinfo {title}
  {Exact stochastic simulation of coupled chemical reactions},}\ }\href
  {\doibase 10.1021/j100540a008} {\bibfield  {journal} {\bibinfo  {journal}
  {The Journal of Physical Chemistry}\ }\textbf {\bibinfo {volume} {81}},\
  \bibinfo {pages} {2340--2361} (\bibinfo {year} {1977})},\ \Eprint
  {http://arxiv.org/abs/http://dx.doi.org/10.1021/j100540a008}
  {http://dx.doi.org/10.1021/j100540a008} \BibitemShut {NoStop}%
\bibitem [{\citenamefont {Bortz}\ \emph {et~al.}(1975)\citenamefont {Bortz},
  \citenamefont {Kalos},\ and\ \citenamefont {Lebowitz}}]{Bortz1975}%
  \BibitemOpen
  \bibfield  {author} {\bibinfo {author} {\bibfnamefont {A.B.}\ \bibnamefont
  {Bortz}}, \bibinfo {author} {\bibfnamefont {M.H.}\ \bibnamefont {Kalos}}, \
  and\ \bibinfo {author} {\bibfnamefont {J.L.}\ \bibnamefont {Lebowitz}},\
  }\bibfield  {title} {\enquote {\bibinfo {title} {A new algorithm for monte
  carlo simulation of ising spin systems},}\ }\href {\doibase
  https://doi.org/10.1016/0021-9991(75)90060-1} {\bibfield  {journal} {\bibinfo
   {journal} {Journal of Computational Physics}\ }\textbf {\bibinfo {volume}
  {17}},\ \bibinfo {pages} {10 -- 18} (\bibinfo {year} {1975})}\BibitemShut
  {NoStop}%
\bibitem [{\citenamefont {Prados}\ \emph {et~al.}(1997)\citenamefont {Prados},
  \citenamefont {Brey},\ and\ \citenamefont {S{\'a}nchez-Rey}}]{Prados1997}%
  \BibitemOpen
  \bibfield  {author} {\bibinfo {author} {\bibfnamefont {A.}~\bibnamefont
  {Prados}}, \bibinfo {author} {\bibfnamefont {J.~J.}\ \bibnamefont {Brey}}, \
  and\ \bibinfo {author} {\bibfnamefont {B.}~\bibnamefont {S{\'a}nchez-Rey}},\
  }\bibfield  {title} {\enquote {\bibinfo {title} {A dynamical monte carlo
  algorithm for master equations with time-dependent transition rates},}\
  }\href {\doibase 10.1007/BF02765541} {\bibfield  {journal} {\bibinfo
  {journal} {Journal of Statistical Physics}\ }\textbf {\bibinfo {volume}
  {89}},\ \bibinfo {pages} {709--734} (\bibinfo {year} {1997})}\BibitemShut
  {NoStop}%
\bibitem [{\citenamefont {Hellier}(2018)}]{jHellGitRepo}%
  \BibitemOpen
  \bibfield  {author} {\bibinfo {author} {\bibfnamefont {Joshua}\ \bibnamefont
  {Hellier}},\ }\bibfield  {title} {\enquote {\bibinfo {title}
  {joshuahellier/phdstuff: Codes for sticky particles in 1d and 2d},}\ }\href
  {\doibase 10.5281/zenodo.1162818} {\  (\bibinfo {year} {2018}),\
  10.5281/zenodo.1162818}\BibitemShut {NoStop}%
\bibitem [{\citenamefont {Robert}(2015)}]{metHastAlg}%
  \BibitemOpen
  \bibfield  {author} {\bibinfo {author} {\bibfnamefont {Christian~P.}\
  \bibnamefont {Robert}},\ }\enquote {\bibinfo {title} {The metropolis-hastings
  algorithm},}\ in\ \href {\doibase 10.1002/9781118445112.stat07834} {\emph
  {\bibinfo {booktitle} {Wiley StatsRef: Statistics Reference Online}}}\
  (\bibinfo  {publisher} {American Cancer Society},\ \bibinfo {year} {2015})\
  pp.\ \bibinfo {pages} {1--15}\BibitemShut {NoStop}%
\bibitem [{\citenamefont {Blythe}\ and\ \citenamefont
  {Evans}(2007)}]{blytheEvans2007}%
  \BibitemOpen
  \bibfield  {author} {\bibinfo {author} {\bibfnamefont {R~A}\ \bibnamefont
  {Blythe}}\ and\ \bibinfo {author} {\bibfnamefont {M~R}\ \bibnamefont
  {Evans}},\ }\bibfield  {title} {\enquote {\bibinfo {title} {Nonequilibrium
  steady states of matrix-product form: a solver's guide},}\ }\href
  {http://stacks.iop.org/1751-8121/40/i=46/a=R01} {\bibfield  {journal}
  {\bibinfo  {journal} {Journal of Physics A: Mathematical and Theoretical}\
  }\textbf {\bibinfo {volume} {40}},\ \bibinfo {pages} {R333} (\bibinfo {year}
  {2007})}\BibitemShut {NoStop}%
\bibitem [{\citenamefont {Appert-Rolland}\ \emph {et~al.}(2015)\citenamefont
  {Appert-Rolland}, \citenamefont {Ebbinghaus},\ and\ \citenamefont
  {Santen}}]{AppertRolland2015}%
  \BibitemOpen
  \bibfield  {author} {\bibinfo {author} {\bibfnamefont {C{\'e}cile}\
  \bibnamefont {Appert-Rolland}}, \bibinfo {author} {\bibfnamefont
  {Maximilian}\ \bibnamefont {Ebbinghaus}}, \ and\ \bibinfo {author}
  {\bibfnamefont {Ludger}\ \bibnamefont {Santen}},\ }\bibfield  {title}
  {\enquote {\bibinfo {title} {Intracellular transport driven by cytoskeletal
  motors: General mechanisms and defects},}\ \ }(\bibinfo {year}
  {2015})\BibitemShut {NoStop}%
\end{thebibliography}

%

\end{document}